\documentclass[11pt]{article}
\pdfoutput=1
\usepackage{jheparxiv}

\usepackage{amsmath}
\usepackage{mathtools}
\usepackage{amssymb}
\usepackage{slashed}
\usepackage{color}
\usepackage{xcolor}
\usepackage{hyperref}
\hypersetup{colorlinks=true,citecolor=blue} \usepackage{braket}
\usepackage{comment}
\usepackage{scalerel}
\usepackage{tikz}
\usepackage{subcaption}
\usetikzlibrary{decorations.pathmorphing,decorations.markings}
\tikzset{snake it/.style={decorate, decoration=snake}}

\newcommand{\be}{\begin{equation}}
\newcommand{\ee}{\end{equation}}
\newcommand{\bi}{\begin{enumerate}}
\newcommand{\ei}{\end{enumerate}}
\newcommand{\ud}{{\mathrm{d}}}
\hfuzz=\maxdimen \tolerance=10000
\vfuzz=\maxdimen \tolerance=10000
\newcommand{\LCm}{{\scriptscriptstyle -}}
\newcommand{\LCp}{{\scriptscriptstyle +}}
\newcommand{\LCpm}{{\scriptscriptstyle \pm}}

\newcommand{\LCperp}{{\scriptscriptstyle \perp}}
\newcommand{\nroots}{\jbar{N}}
\newcommand*\jbar[1]{%
  \hbox{%
    \vbox{%
      \hrule height 0.1pt 
      \kern0.5ex
      \hbox{%
        \kern-0.15em
        \ensuremath{#1}%
        \kern-0.02em
      }%
    }%
  }%
} 

\renewcommand{\j}{\theta}

\newcommand{\e}{\mathrm{e}}

\begin{document}

\title{All-multiplicity  amplitudes in impulsive PP-waves from the worldline formalism}

\author[a]{Patrick Copinger,}
\emailAdd{patrick.copinger@plymouth.ac.uk}
\author[a]{James P.~Edwards,}
\emailAdd{james.p.edwards@plymouth.ac.uk}
\author[b]{Anton Ilderton}
\emailAdd{anton.ilderton@ed.ac.uk}
\author[b]{and Karthik Rajeev}
\emailAdd{karthik.rajeev@ed.ac.uk}

\affiliation[a]{Centre for Mathematical Sciences, University of Plymouth, Plymouth, PL4 8AA, UK}

\affiliation[b]{Higgs Centre, School of Physics and Astronomy, University of Edinburgh, EH9 3FD, UK}

\abstract{
We use the worldline formalism to derive Bern-Kosower type Master Formulae for the tree-level scattering of a charged particle and an arbitrary number of photons on impulsive PP-waves, where the coupling of the PP-wave to matter is treated fully non-perturbatively. We show that, in a certain kinematic regime characterised by a semi-classical positive energy condition, both off-shell currents and scattering amplitudes exhibit two novel factorisation structures. First, they may be written as currents in vacuum but with a single additional photon, averaged over the momentum of that 
photon. This converts the all-orders interaction with the PP-wave into a single effective interaction. Second, the currents and amplitudes may be written as a weighted average of the corresponding quantities in an impulsive plane wave background, with the average taken over all possible field strengths of the plane wave. This generalises a known single-photon result to arbitrary multiplicity.}

\maketitle

\section{Introduction}
%
It is well known that the elastic scattering amplitude of two charged particles
at high centre of mass energy and low momentum transfer is
well-approximated by the
resummation of all ladder and cross-ladder diagrams~\cite{Abarbanel:1969ek,Levy:1969cr,Brezin:1970,Tiktopoulos:1971hi,Eichten:1971kd}, dominated by mediation of the highest spin boson~\cite{tHooft:1987vrq}. The resulting `eikonal' amplitude can, for the gravitational scattering of massive scalar particles, be recast as 1$\rightarrow 1$ scattering of a massless scalar on the background of an exact `shockwave' solution of the vacuum Einstein equations~\cite{Aichelburg:1970dh}, now interpreted as the space-time generated by another ultra-relativistic particle~\cite{Bonnor:1969mfs} (c.f.~\cite{Cristofoli:2020hnk} for an amplitudes-based derivation of the metric).  This correspondence has been exploited to study e.g.~transplankian gravitational scattering~\cite{Verlinde:1991iu,Kabat:1992tb,Lodone:2009qe}. The connection between QFT in more general spacetimes and the eikonal limit was recently explored in~\cite{Adamo:2021rfq}. The analogous story in QED is that eikonal $2\to 2$ scattering may be recast as $1\to1$ scattering on the shockwave which is the ultraboosted Coulomb field of a point charge~\cite{Bonnor:1969rb,Jackiw:1991ck}.

The example of eikonal scattering is just one of many scenarios in the paradigm of \,`QFT in a background field'~\cite{Furry:1951zz,DeWitt:1967ub,tHooft:1975uxh,Abbott:1981ke}, which effectively bypasses explicit evaluations of all-order summations in certain limits of the process of interest. This framework is particularly relevant when the coupling of particles to the background is large and must be treated non-perturbatively. Such `strong' backgrounds can arise in e.g.~gauge theories in magnetic fields~\cite{Miransky:2015ava,Hattori:2016emy,Hattori:2023egw}, in magnetars~\cite{Turolla:2015mwa,Kaspi:2017fwg,Kim:2021kif} and in other astrophysical environments~\cite{Hu:2019nyw,Harding:2006qn,Baring:2008aw, Baring:2000cr}. Terrestrially, strong electromagnetic fields may be produced by high-intensity lasers~\cite{Strickland:1985gxr}, and both nonlinear and non-perturbative aspects of `strong field QED' are the target of upcoming experiments at high intensity laser facilities~\cite{Abramowicz:2021zja,Karbstein:2021ldz,Clarke:2022rbd}. For reviews see~\cite{Gonoskov:2021hwf,Fedotov:2022ely}.

An important advantage of background-field methods is that they provide efficient access to non-linear and non-perturbative aspects of theories and the processes they model. This is enabled by working in the Furry picture~\cite{Furry:1951zz, Seipt:2017ckc}, whereby interactions with the fixed background field are treated to all orders through suitably background-dressed propagators and asymptotic wavefunctions, while interactions between quantised particles continue to be treated perturbatively. However, this approach is often hindered by the fact that such calculations are significantly more challenging than the corresponding calculations in vacuum. Even at tree level and at low multiplicity (e.g.~$1\to2$ on the background), scattering amplitudes can exhibit arbitrary complexity, i.e.~they are no longer rational functions of kinematic data.
For e.g.~QED in plane wave backgrounds, modelling an intense laser, the state of the art using standard diagrammatic methods is, at tree level, only \emph{four} points, and only recently have attempts been made at deriving formulae suitable for higher multiplicity~\cite{Edwards:2020npu,103,Edwards:2021vhg,Schubert:2023gsl} (see also~\cite{Edwards:2018vjd,Ahmadiniaz:2023jwd} for all-multiplicity results in homogeneous backgrounds). This is a serious issue since upcoming experiments demand input from both higher loop and higher multiplicity scattering processes.

{Motivated by the need to go to higher multiplicity, we consider in this paper QED scattering in a particular class of backgrounds which is amenable to analytic progress. Our chosen fields are inspired by the eikonal results above, and are natural generalisations of the shockwave obtained by ultra-boosting \emph{arbitrary} source distributions; these are called `impulsive PP-waves'. As these are ultra-boosted they are compactly supported on null rays, and this locality provides significant simplifications in scattering processes, compared to other commonly studied backgrounds such as constant fields and plane waves. We allow the impulsive background to be arbitrarily strong, and hence we will treat it non-perturbatively in the Furry picture throughout.}
    
We will employ here the first quantised approach to QFT known as the worldline formalism~\cite{UsRep,ChrisRev}.  Crucially this will allow us to compute \emph{all-multiplicity} tree-level amplitudes with one matter line and $N$ photon legs, for arbitrary $N$, in an arbitrary PP-wave background. The majority of our results are obtained without necessarily calculating explicit expressions for the scattering amplitudes, but rather by manipulations under the worldline path integral. {Conveniently, this formalism combines contributions from what would be multiple Feynman diagrams into one calculation (summing over permutations of the external photons absorbed or emitted from the line).} This allows us to make more general statements about arbitrary multiplicity amplitudes, and to relate amplitudes in PP-waves to amplitudes calculated using worldline techniques both in vacuum \cite{Ahmadiniaz:2020wlm, Ahmadiniaz:2021gsd} and in plane wave backgrounds~\cite{Edwards:2021vhg, Copinger:2023ctz}. Specifically, we will show that the all-orders interaction with an impulsive PP-wave can be recast as an effective single-photon interaction in vacuum. We will also show that amplitudes in impulsive PP-wave backgrounds can be obtained from those in (impulsive) plane wave backgrounds by averaging over the field strength of the latter -- in effect, the PP wave acts as a `stochastic' plane wave. This generalises the single-photon-emission result obtained in~\cite{Adamo:2021jxz} to arbitrary photon multiplicity.

Although the worldline formalism has had considerable success when applied to processes in homogeneous~\cite{AFFLECK1982509,Fradkin:1991zq,Adler:1996cja, Reuter:1996zm,shaisultanov1996string,Srinivasan:1998ty,Schubert:2001he,Edwards:2017bte,Ahmadiniaz:2017rrk} and plane wave~\cite{Ilderton:2016qpj,Copinger:2023ctz,  Schubert:2023gsl, Edwards:2021vhg} backgrounds, there are far fewer examples where it has been used to study processes in shockwave backgrounds -- see though~\cite{Tarasov:2019rfp} for an application to deep inelastic scattering, where the worldline path integral relevant to the one-loop effective action was evaluated analytically within a certain eikonal approximation. Here we will be able to treat the path integral exactly, albeit at tree-level for the time being, with the anticipation of generalising our approach to loop corrections at a later stage. 

This paper is organised as follows. In Sec.~\ref{sect:momentum-propagator} we first review the definition and evaluation of the `$N$-photon dressed propagator', that is the correlation function of 2 massive particles and $N$ photons, the latter being LSZ truncated. This is the central object we will study. We also review the off-shell currents and amplitudes obtained by further LSZ reduction of the massive particles in the dressed propagator. The emphasis here is on backgrounds for which the worldline path integrals evaluate \emph{exactly} to its semi-classical value, as applies to impulsive PP-waves. We work at tree level throughout. Sec.~\ref{sec:scalar} details our two main results in the case of scalar QED, and in Sec.~\ref{sec:spinor} we extend these results to QED proper. We conclude in Sec.~\ref{sect:conclusions}.

\subsection*{Notation and conventions}
We work in Minkowski space throughout, mostly minus signature, and largely in lightfront coordinates $x^\LCpm=x^0\pm x^3$, $x^\LCperp=(x^1,x^2)$. To tidy factors of $2\pi$ in integral measures and delta functions we write ${\hat \ud}=\ud/(2\pi)$ and ${\hat \delta}=(2\pi)\delta$. 

\section{$N$-photon dressed propagator and off-shell current}\label{sect:momentum-propagator}
%
Starting with the correlation functions of 2 massive particles and $N$ photons, the `$N$-photon dressed propagator' is obtained by LSZ truncating the photon legs, although \emph{not} necessarily taking the photon momenta to be on-shell. These are natural objects to consider in the worldline approach, and they are easily connected to other familiar objects; applying LSZ to the massive particles yields `off-shell currents', and if one additionally sets the  photon momenta on-shell one recovers 2-particle, $N$-photon scattering amplitudes.

Conventionally, the momentum space dressed propagator is obtained from Fourier transforming its position space representation, but we expand here on a method outlined in~\cite{Copinger:2023ctz} for a more direct evaluation based on the worldline path integral representation of the propagator.  For the purpose of illustration and sake of brevity, we focus first on the scalar QED case; the extension to the spinor case is then straightforward.

The worldline formalism~\cite{Feyn1,Feyn2,Strass1}
reformulates field theory in terms of path integrals for a fixed number of relativistic point particles coupled to a suitable gauge field (for reviews see \cite{UsRep, ChrisRev}). The worldline action weighting this path integral encodes the emission and absorption of an arbitrary number of photons along the particle trajectories. In this framework, the dressed momentum-space propagator for scalar QED, as presented in \cite{Copinger:2023ctz}, is 
\begin{align}\label{propagator_mom_PI}
    \mathcal{D}^{p'p}_{N} = 
    (-ie)^{N}\!
    \int_{0}^{\infty}\!\ud T 
    \int_{0}^{T}\prod_{i=1}^{N} \ud\tau_i
\int\mathcal{D}x(\tau)\, \e^{iS_{\mathcal{J}}[x(\tau);A]}\Big|_{\textrm{lin.}\,\varepsilon}\;,
\end{align}
where the worldine action $S_{\mathcal{J}}$ takes the form
\be
\label{def:SJ}
\begin{split}
   S_{\mathcal{J}}[x;A] = {p'\cdot x(T) - p\cdot x(0)} -\int_{0}^{T}\!\ud\tau\,
   \Big[
   \frac{1}{4}\dot{x}^2(\tau)+m^2+eA(x(\tau))\cdot\dot{x}(\tau)+{\mathcal{J}_\mu x^\mu(\tau)}
   \Big]
   \,,
\end{split}
\ee
in which the first two terms, together with {\textit{free}} boundary conditions in \eqref{propagator_mom_PI}, implement the transformation from position space to Fourier space, the external current $\mathcal{J}$ is
\begin{align}\label{ext_current}
    \mathcal{J}^{\mu}=i\sum_{i=1}^{N}\big(ik_{i}^\mu - \varepsilon_{i}^\mu\frac{\ud}{\ud\tau}\big)\delta(\tau-\tau_{i})\,,
\end{align}
and `$\textrm{lin.}\,\varepsilon$' is shorthand for the instruction to take only the term linear in (each of the) $\varepsilon_i$.
In \cite{Copinger:2023ctz}, we used this representation to study the dressed propagator, and the associated 2-massive-particle, $N$-photon scattering amplitudes, in general plane-wave backgrounds. Here we present an alternate `mixed' version of \eqref{propagator_mom_PI} which is convenient for applications. The only difference is that in this new representation, one keeps the Fourier transform with respect to the 
initial point explicit; more precisely, we write
\begin{align}\label{propagator_mom_PI_mixed}
    \mathcal{D}^{p'p}_{N} =(-ie)^{N}\!\int \! \ud^{4}x    
    \int_{0}^{\infty}\!\ud T \,
    \int_{0}^{T}\prod_{i=1}^{N} \ud\tau_i\,
\int_{x(0)=x}\hspace{-1.5em}\mathcal{D}x(\tau)\, \e^{iS_{\mathcal{J}}[x(\tau);A]}\Big\rvert_{\textrm{lin.}\,\varepsilon}\;.
\end{align}
To understand the utility of this representation, let us first consider a general background and calculate the worldline path integral in (\ref{propagator_mom_PI_mixed}) \emph{in the semiclassical limit}~\cite{Dunne:2005sx,Dunne:2006st} (something we will relax later). Doing so returns, as is standard, the exponential of the classical action, that is the action evaluated on a solution $x_{\text{cl}}^\mu$ of the equations of motion, multiplied by a fluctuation determinant~$\mathcal{N}$:
\begin{align}\label{propagator_mom_PI_semi}
    \mathcal{D}^{p'p}_{N} =(-ie)^{N}\!\int \! \ud^{4}x\,
    \int_{0}^{\infty}\!\ud T \,
    \int_{0}^{T}\prod_{i=1}^{N} \ud\tau_i\,
\mathcal{N}(p',x,T;\mathcal{J})\, \e^{iS_{\mathcal{J}}[x_{\text{cl}}(\tau);A]}\Big|_{\textrm{lin.}\,\varepsilon}\;.
\end{align}
The equations of motion determining $x^\mu_{\text{cl}}$ are found by varying the action (\ref{def:SJ}); this yields
\be
\label{vary:SJ}
\begin{split}
   \delta S_{\mathcal{J}}[x;A] = \big[ p' -\tfrac12 \dot{x}(T) &- e A(x(T))\big]\cdot \delta x(T) - \big[p - \tfrac12 \dot{x}(0) - e A(x(0))\big] \cdot \delta x(0) \\
   &+\int_{0}^{T}\!\ud\tau\, \delta x^\mu \Big( \frac12 \ddot{x}_\mu  -F_{\mu\nu}(x){\dot x}^\nu  -\mathcal{J}_\mu\Big) \,.
\end{split}
\ee
Consider the terms in square brackets; on the lower boundary the variation vanishes by the Dirichlet condition on the path integral, whereas the variation on the upper boundary vanishes provided we choose the classical path to obey mixed, or Robin, boundary conditions
\be
    \tfrac12 \dot{x}_\mu(T)+ eA_\mu(x(T)) = p'_\mu \;.
\ee
Using this, some algebra and integration by parts (keeping track of boundary terms) reduce the classical action to\footnote{Note that $\dot{x}_{\text{cl}}^2$ appears with a positive sign, as opposed to how it appears in the definition of $S_{\mathcal{J}}[x;A]$. The equations of motion and boundary term conspire to induce this change of sign of the kinetic term.},
\be\begin{split}\label{Scl}
    S_{\mathcal{J}}[x_{\text{cl}};A]
    &=
    (p'+K-p)\cdot x
    + \int_{0}^{T}\!\ud\tau \,
    \Big(\frac{1}{4}\dot{x}_{\text{cl}}^2-m^2\Big)\\
    & +e\int_{0}^{T}\!\ud\tau\, \dot{x}_{\text{cl}}^{\mu}(\tau)\int_{\tau}^{T}\!\ud\tau'\, \dot{x}_{\text{cl}}^{\nu}(\tau')\partial_{\mu}A_{\nu}(x_{\text{cl}}(\tau')) \,,
\end{split}
\ee
in which we have defined $K=\sum_{i}^{N}k_i$ as the sum of external photon momenta. Here the usefulness of our representation starts to become apparent; information about momentum conservation (or lack thereof) is conveniently packaged into the first term of (\ref{Scl}) when considered with the integral over the initial point, while the pole at the outgoing on-shell momentum is found to be nicely encapsulated in the second term. This makes LSZ reduction of the outgoing leg particularly simple. 

{From these general considerations we now restrict to a class of backgrounds for which the semiclassical approximation to the path integral in \eqref{propagator_mom_PI_mixed} is \emph{exact}. This class contains (at least) constant fields and plane waves, as well as the vacuum limit. Furthermore, the fluctuation determinant $\mathcal{N}$ is unity in both the vacuum and plane wave cases (contrast this with the homogeneous field, where this determinant {can contain} important physical information~\cite{AFFLECK1982509}).

To illustrate this, we consider here the case of a plane wave background in the gauge \newline $eA_{\mu}=\delta^{\LCperp}_{\mu}a_{\perp}(n\cdot x)$. The classical action then simplifies further and may be written
\begin{align}\label{S-cl-pw}
    S_{\mathcal{J}}[x_{\text{cl}};A]= \left(p'+K-p\right)\cdot x  + \int_{0}^{T}\!\ud\tau\, \left(\frac{1}{4}v^2-m^2\right)\,,
\end{align}
where $v=2p'+ 2\sum_{i=1}^{N}\left[k_i\Theta(\tau_i-\tau)-i\varepsilon_{i}\delta(\tau-\tau_i)\right]-2eA(x_{\text{cl}})$. Upon expanding $v^2$ and integrating, we can see how the pole of the propagator as the outgoing momentum $p'$ goes on-shell emerges from the term $(p'^2-m^2)T$. 

The explicit form of the dressed propagator, both in a plane wave background and in vacuum, will be needed later, so we give both here. For the plane wave case we have in lightfront gauge $\varepsilon_{i}^\LCp=0$ \cite{Copinger:2023ctz},
\begin{align}\label{master-pw}
    \mathcal{D}^{(\textrm{pw})p'p}_{N} =(-ie)^{N}\!\int \! \ud^{4}x\!
    \int_{0}^{\infty}\!\ud T\, & \e^{i({\tilde p}'^2-m^2) T}
    \int_{0}^{T}\prod_{i=1}^{N}\! \ud\tau_i \\ \nonumber
    \e^{i({\tilde p}'+K-p)\cdot x} & \e^{ i(2{\tilde p}'+K)\cdot g-i\sum_{i,j=1}^{N}\Big[\frac{1}{2}|\tau_i-\tau_j|k_i\cdot k_j-i\,\,\textrm{sgn}(\tau_i-\tau_j)\varepsilon_i\cdot k_j+\varepsilon_i\cdot\varepsilon_j\delta(\tau_i-\tau_j)\Big]}\\\nonumber
    &\qquad \qquad \quad \!\!\e^{-\int_{0}^{T}(2{\tilde p}' \cdot a(\tau)-a^2(\tau))\ud\tau-2\sum_{i}^{N}\left[\int_{0}^{\tau_i}k_{i}\cdot a(\tau)\ud\tau-i\varepsilon_i\cdot a(\tau_i)\right]}\Big|_{\textrm{lin.}\,\varepsilon}\;,
\end{align}
in which ${\tilde p}' = p'+ a^{\infty}$ accounts for the memory effect~\cite{Kibble:1965zza,Dinu:2012tj,Cristofoli:2022phh}, we have introduced
\begin{equation}
    a_{\mu}(\tau) \equiv 
    a(x^\LCp_{\rm cl}(\tau))\equiv 
    a_\mu\Bigl(x^\LCp + g^{\LCp}
    +(p'+p)^\LCp\tau-\sum_{i=1}^Nk_{i}^{\LCp}|\tau-\tau_i|\Bigr)\,,
\end{equation}
and $g$ is defined by
\be
   g \equiv g(\{\tau_i\}):= \sum_{j=1}^{N}(k_j\tau_j-i\varepsilon_j) \;.
\ee
In the vacuum limit, many of the exponential terms vanish, and the $\ud^4x$ integral can be performed to recover the conservation of four-momentum, thus:
\begin{align}
\label{master-vac}
    D^{p'p}_{N} =(-ie)^{N}&{\hat\delta}^4(p'+K-p)
    \int_{0}^{\infty}\!\ud T\,
    \e^{i(p'^2-m^2) T}
    \int_{0}^{T}\prod_{i=1}^{N}\! \ud\tau_i \\ \nonumber
    & \e^{ i(2p'+K)\cdot g-i\sum_{i,j=1}^{N}\Big[\frac{1}{2}|\tau_i-\tau_j|k_i\cdot k_j-i\,\,\textrm{sgn}(\tau_i-\tau_j)\varepsilon_i\cdot k_j+\varepsilon_i\cdot\varepsilon_j\delta(\tau_i-\tau_j)\Big]}\bigg|_{\textrm{lin.}\,\varepsilon}\;.
\end{align}
In order to compactify notation, we will henceforth abbreviate 
\be
    \int_N := (-ie)^{N}\int \ud^4x\!
    \int_{0}^{\infty}\!\ud T
    \int_{0}^{T}\prod_{j=1}^{N} \ud\tau_j \;,
\ee
as this collection of integrals appears throughout the paper.
\begin{figure}
     \centering
     \begin{tikzpicture}
    \begin{scope}[decoration={
    markings,
    mark=at position 0.5 with {\arrow{>}}}
    ] 
            \draw[line width=0.2mm,double,postaction={decorate}] (-1.5,0) -- (-.5,0);
            \draw (-1.4,-.3) node {$p$};
            \draw[line width=0.2mm,double,postaction={decorate}] (.5,0) -- (1.5,0);
            \draw (1.4,-.3) node {$p'$};
            \draw[draw,fill=lightgray] (0,0) circle (0.5) node {$\mathcal{D}^{p'p}_{N}$};
            \draw[draw=blue, snake it] (-0.433013, 0.25) -- (-1.29904, 0.75);
            \draw (-1.47224, 0.85) node {$k_1$};
            \draw[draw=blue, snake it] (-0.25, 0.433013) -- (-0.75, 1.29904);
            \draw (-0.85, 1.5) node {$k_2$};
            \filldraw[black] (0.5, 0.866025) circle (.5pt);
            \filldraw[black] (0.258819, 0.965926) circle (.5pt);
            \filldraw[black] (0, 1) circle (.5pt);
            \draw[draw=blue, snake it] (0.433013, 0.25) -- (1.29904, 0.75);
            \draw (1.47224, 0.9) node {$k_N$};
            \draw (1.8,-.8) --  (1.8,1);
            \draw (3,-.75) node {$k_i^2,p'^2,p^2\!\in\!\mathbb{R}$};
            \draw[line width=0.2mm, ->] (4.2,0) -- (6.2,0);
            \draw (5.2,.3) node {LSZ on $p'$, $p$};
            \draw[line width=0.2mm,double,postaction={decorate}] (-1.5+8.2,0) -- (-.5+8.2,0);
            \draw (-1.4+8.2,-.3) node {$p$};
            \draw[line width=0.2mm,double,postaction={decorate}] (.5+8.2,0) -- (1.5+8.2,0);
            \draw (1.4+8.2,-.3) node {$p'$};
            \draw[draw,fill=lightgray] (8.2,0) circle (0.5) node {$\mathcal{A}^{p'p}_{N}$};
            \draw[draw=blue, snake it] (-0.433013+8.2, 0.25) -- (-1.29904+8.2, 0.75);
            \draw (-1.47224+8.2, 0.85) node {$k_1$};
            \draw[draw=blue, snake it] (-0.25+8.2, 0.433013) -- (-0.75+8.2, 1.29904);
            \draw (-0.85+8.2, 1.5) node {$k_2$};
            \filldraw[black] (0.5+8.2, 0.866025) circle (.5pt);
            \filldraw[black] (0.258819+8.2, 0.965926) circle (.5pt);
            \filldraw[black] (8.2, 1) circle (.5pt);
            \draw[draw=blue, snake it] (0.433013+8.2, 0.25) -- (1.29904+8.2, 0.75);
            \draw (1.47224+8.2, 0.9) node {$k_N$};
            \draw (1.8+8.2,-.8) --  (1.8+8.2,1);
            \draw (3+8.8,-.75) node {$k_i^2\in\mathbb{R};p'^2,p^2\!=\! m^2$};       
        \end{scope}
\end{tikzpicture}
\caption{{\textit{Left:} the dressed propagator $\mathcal{D}^{p'p}_{N}$. 
The double line represents the non-perturbative treatment of the coupling between charged particles and the background.
\textit{Right:} LSZ reduction on the matter lines in $\mathcal{D}^{p'p}_{N}$ yields the tree-level off-shell current $\mathcal{A}^{p'p}_{N}$, wherein the photon momenta $k_i$ are not necessarily on-shell. One can compute the corresponding tree-level amplitude from $\mathcal{A}^{p'p}_{N}$ by simply setting the photons to be on-shell and transverse, i.e.~$k_i^2= 0 = \varepsilon_{i}\cdot k_{i}$.}} \label{fig:DtoA}
\end{figure}
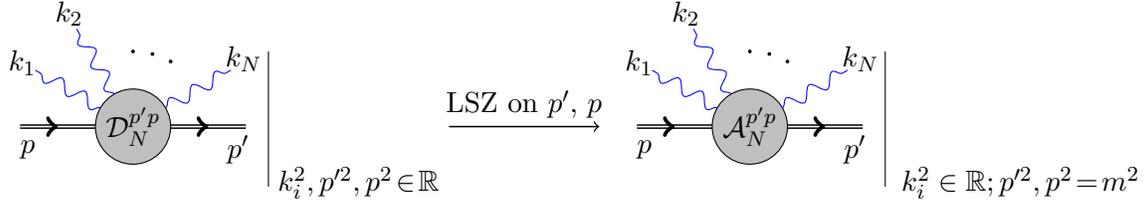 

\subsection{Off-shell currents}
\label{secAmputation}
We now move to the discussion of the off-shell current $\mathcal{A}^{p'p}_{N}$, defined as the result of LSZ reduction applied to the matter lines in the dressed propagator $\mathcal{D}_{N}^{p'p}$, see Fig.(\ref{fig:DtoA}). Consequently, while all legs in $\mathcal{A}^{p'p}_{N}$ are truncated, only the matter momenta are necessarily on-shell; when evaluated, in addition, at on-shell photon momenta, the off-shell current reduces to the corresponding tree-level scattering amplitude.

As well as being an intermediate step in the computation of tree-level scattering amplitudes from the dressed propagator, the off-shell currents have several interesting applications in and of themselves. In their pioneering work~\cite{Berends:1987me}, Berends and Giele utilised off-shell currents to develop a recursive method of calculating tree-level multi-gluon scattering amplitudes.  Higher-loop corrections may also be efficiently generated by stitching pairs of off-shell photon legs together, see for example~\cite{Schmidt:1994aq,Schubert:2001he,Bastianelli:2014bfa,Ahmadiniaz:2020jgo,Ahmadiniaz:2023vrk}). The classical limit of off-shell currents has also received significant attention in the study of gravitational dynamics~\cite{Goldberger:2016iau,Goldberger:2017frp,Goldberger:2017ogt,Shen:2018ebu}. For a recent review of the classical limit of various off-shell currents and their application, see \cite{Comberiati:2022ldk}.

It is useful to review the LSZ truncation of the dressed propagator in its worldline representation, both in vacuum and plane wave backgrounds. In vacuum, reduction of the matter lines can be achieved by the following `shorthand' of operations on the master formula for ${D}^{p'p}_{N}$~\cite{Mogull:2020sak}:
\begin{enumerate}
    \item[(i)] delete the integral $\int \ud T$,\label{rules1}
    \item[(ii)] insert a factor of $\delta\Big(\sum^{N}_{i=1}\tau_i/N\Big)$,
    \item[(iii)] change the range of the $\ud\tau_i$ integrals to $\mathbb{R}$.
\end{enumerate}  
{These rules turn \eqref{master-vac} into 
\begin{align}\label{master-vac-amplitude}
    \mathcal{A}^{p'p}_{N}=&(-ie)^N
    \hat{\delta}^4({\tilde p}'+K-p)
    \int_{-\infty}^{\infty}\prod_{i=1}^{N} \ud\tau_i\, 
    \delta\bigg(\sum_{j=1}^N\frac{\tau_j}{N}\bigg)
    \nonumber \\
 & \e^{i({\tilde p}' +p)\cdot g-i\sum_{i,j=1}^{N}\bigl(\frac{1}{2}|\tau_i-\tau_j|k_i \cdot k_j-i\,\mathrm{sgn}(\tau_i-\tau_j)\varepsilon_i \cdot k_j+\delta(\tau_i-\tau_j)\varepsilon_i \cdot \varepsilon_j\bigr)}
 \Big|_{\textrm{lin. }\varepsilon} \;,
 \end{align}}
In the plane wave case, the analogous prescription to effect LSZ reduction of  $\mathcal{D}^{(\text{pw})p'p}_{N}$ comprises rules (i)-(iii) and in addition~\cite{Copinger:2023ctz}:
\begin{enumerate}
    \item[(iv)] introduce the factor $\exp{(-i\int_{\infty}^{0}p'\cdot a(\infty)}\ud\tau)$, to cancel spurious divergent factors induced by rule (iii).
\end{enumerate}
This turns \eqref{master-pw} into
\begin{align}\label{master-pw-amplitude}
 	\mathcal{A}^{p'p}_{N}=&(-ie)^N
  {\hat \delta}_{\perp,-}({\tilde p}'+K-p)\int_{-\infty}^{\infty}\!\ud x^{\LCp}
  \e^{i(K+p'-p)_\LCp x^{\LCp}}
 \int_{-\infty}^{\infty}\prod_{i=1}^{N} \ud\tau_i\, 
 \delta\bigg(\sum_{j=1}^N\frac{\tau_j}{N}\bigg)
  \nonumber \\
 &
 \e^{-i\int_{-\infty}^{0}
 [2\tilde{p}' \cdot a(\tau)-a^{2}(\tau)]\ud\tau
 -i\int_{0}^{\infty}\!
 [2p'\cdot\delta a(\tau)-\delta a^{2}(\tau)]\ud\tau
 -2i\sum_{i=1}^{N}
 [\int_{-\infty}^{\tau_i}k_{i} \cdot a(\tau)\ud\tau-i\varepsilon_i\cdot a(\tau_i)]}\nonumber \\
 & \e^{i({\tilde p}' +p)\cdot g-i\sum_{i,j=1}^{N}\bigl(\frac{1}{2}|\tau_i-\tau_j|k_i \cdot k_j-i\,\mathrm{sgn}(\tau_i-\tau_j)\varepsilon_i \cdot k_j+\delta(\tau_i-\tau_j)\varepsilon_i \cdot \varepsilon_j\bigr)}
 \Big|_{\textrm{lin. }\varepsilon} \;.
 \end{align}
These `rules' indicate that while the relevant domain for the auxiliary particle worldlines in  $\mathcal{D}^{p'p}_{N}$ is $\tau\in(0,T)$, that for $\mathcal{A}^{p'p}_{N}$ is the entire real-line $\mathbb{R}$. This will be particularly important below. It would be interesting to see how these prescriptions generalise to other background fields.

We believe that \eqref{propagator_mom_PI_semi}, with the classical action in the form \eqref{Scl}, is a convenient starting point for studying off-shell currents and scattering amplitudes in a wide class of external backgrounds. In the following sections, we show how this general formalism, along with its application in the specific case of plane wave backgrounds~\cite{Copinger:2023ctz}, can be used to compute the master formula for scattering amplitudes in a PP-wave background field.

\clearpage

\section{Scalar particles scattering on impulsive PP-waves}\label{sec:scalar}
An impulsive PP-wave may be defined by the potential
\begin{align}\label{PP-wave-def}
    A^{(s)}_{\mu}(x)=-n_{\mu}\delta(n\cdot x)\Phi(x^{\LCperp})\,,
\end{align}
in which $n^2=0$ while $\Phi(x^{\LCperp})$ is an arbitrary function of the `transverse' coordinates $x^\LCperp$; if we go to a frame where $n\cdot x = x^\LCp \equiv x^0+x^3$ then $x^\LCperp=\{x^1,x^2\}$. The field strength and source of the wave are
\be\label{F-J-PP}
    F^{(s)\mu\nu} = \delta(n\cdot x) (n^\mu \partial^\nu -n^\nu \partial^\mu) \Phi \;,
    \qquad
    \quad
    \partial_\mu F^{(s)\mu\nu} = n^{\nu}\delta(n\cdot x)\nabla^2_{\perp}\Phi \;.
\ee
Particular choices of $\Phi(x^\LCperp)$ of interest include the shockwaves (the ultra-boosted Coulomb fields of charged particles), {$\Phi(x^\LCperp)\propto \log(\mu^2|x^{\LCperp}|^2)$ with $\mu$ a constant}, and impulsive plane waves (vacuum solutions) $\Phi(x^\LCperp) = r_\LCperp x^\LCperp$ where $r_\LCperp$ is a constant vector.  We keep $\Phi$ arbitrary throughout (and later we will see how scattering on any impulsive PP-wave, i.e.~any $\Phi$, is related to scattering in the plane wave case).

We now proceed to evaluate the dressed propagator (\ref{propagator_mom_PI}), with the aim of extracting from it the off-shell currents and amplitudes for scalar particles scattering on the PP-wave background (\ref{PP-wave-def}), and emitting/absorbing $N$ photons. We work in lightfront gauge $n\cdot\varepsilon\equiv\varepsilon^\LCp=0$ throughout. (This condition is also obeyed by the background (\ref{PP-wave-def}).) {To perform the path integrals in (\ref{propagator_mom_PI}), we first split the lightfront coordinates into the classical solution in vacuum and a quantum fluctuation, \textit{viz}.,
\begin{equation}
    x^{\pm}(\tau)=x^{\pm}_{\text{cl}}(\tau)+\delta x^{\pm}(\tau)\,, \qquad \ddot{x}^{\LCp}_{\text{cl}}=2\mathcal{J^{\LCp}} \;,
\end{equation}
with $\cal J$ as in (\ref{ext_current}), along with the boundary conditions appropriate the to path integral, $x^{\LCpm}_{\text{cl}}(0)=x^{\LCpm}$ and $\dot{x}^{\LCpm}_{\text{cl}}(T)+2eA^{(s)\LCpm}(x^+(T))=2p'^{\LCpm}$.} {In the PP-wave background~\eqref{PP-wave-def}, the $\mathcal{D}\delta x^{\LCp}\mathcal{D}\delta x^{\LCm}$ integrals can be explicitly evaluated due to a hidden Gaussianity making them semi-classically exact~\cite{Ilderton:2016qpj,Edwards:2021vhg,Edwards:2021uif}; we omit the details of this step since it is virtually the same as that in the plane wave case discussed in \cite{Copinger:2023ctz}.}
The net result of performing the integrals is the replacement $x^{\pm}\rightarrow x_{\rm{cl}}^{\pm}$ in the integrand, so that we arrive at (c.f. (\ref{propagator_mom_PI_semi}))
\begin{align}\label{PI-transverse}
    \mathcal{D}^{(s)p'p}_{N} = {\int_N}\int_{x^{\LCperp}(0)=x^{\LCperp}}\hspace{-1.5em}\mathcal{D}x^{\LCperp}
    \e^{iS_{\mathcal{J}}[(x^{\pm}_{\text{cl}},x^{\LCperp});0]+i\int_{0}^{T}\,\ud\tau\,
    {e(\dot{x}^{\LCp}_{\text{cl}})\delta(x^{\LCp}_{\text{cl}})\Phi(x^{\LCperp})
    }}
    \bigg\rvert_{\textrm{lin.}\,\varepsilon}.
\end{align}
As such, for a general potential $\Phi(x^{\LCperp})$, the path integral over transverse worldline coordinates in the above expression does not seem amenable to exact evaluation. However, motivated by the calculation in~\cite{Adamo:2021jxz}, we adopt the following trick to make progress. First, note that owing to the Dirac-delta function, one can rewrite the field-dependent part of the action as
\begin{align}\label{U1-product}
    \e^{i\int_{0}^{T}\ud\tau\,\dot{x}^{\LCp}_{\text{cl}}\delta(x^{\LCp}_{\text{cl}})\,e\,\Phi(x^{\LCperp}(\tau))}=\prod_{{j}=1}^{\nroots}
    \e^{ie\,\textrm{sgn}(\dot{x}^{\LCp}_{\text{cl}}(t_j))\,\Phi(x^{\LCperp}(t_j))}\,,
\end{align}
where $t_j$, $j=1,2,3,..,\nroots$ are the roots of $x_{\text{cl}}^{\LCp}(\tau)=0$ and `sgn' indicates the sign function.
Therefore, the field-dependence in the path integral is now simply a product of $U(1)$ elements of the form $e^{\pm ie\phi(x^{\LCperp})}$. The trick is to introduce the Fourier transform of these $U(1)$ factors \cite{Adamo:2021jxz, Tarasov:2019rfp}, namely, via
\begin{align}\label{u1-factors}
    \e^{ie\, \textrm{sgn}(\dot{x}^\LCp)\Phi(x^{\LCperp})} =
    \int\! \hat{\ud}^2 r_\LCperp \,
    W(r_{\perp})\, 
    \e^{i\, \textrm{sgn}(\dot{x}^\LCp) r_{\perp} x^{\LCperp}}\,,
\end{align}
and re-write \eqref{PI-transverse} with a linearised action     as
\be\label{PI-transverseX}
\begin{split}
    \mathcal{D}^{(s)p'p}_{N} = {\int_N}\,
\int_{x(0)=x}\!\!\!\!\mathcal{D}x^{\LCperp}(\tau)
\int\!\prod_{j=1}^{\nroots}
    {\hat \ud}^2 r_{j \LCperp}W(r_{j\LCperp})\,
\e^{iS_{\mathcal{J}}[(x^{\LCpm}_{\text{cl}},x^{\LCperp});0]+\,\textrm{sgn}(\dot{x}^{\LCp}(t_j))r_{j\LCperp}x^{\LCperp}(t_j)}\Big\rvert_{\textrm{lin.}\,\varepsilon}\!\!\;.
\end{split}
\ee
The transverse path integral in \eqref{PI-transverseX} is now effectively Gaussian;
whether we choose to evaluate it explicitly or not will lead us to two different representations of the dressed propagator, exposing different structures, as we detail in the following two subsections. 

\subsection{Effective vacuum current}
%
We now compute the $\mathcal{D}x^{\LCperp}$ integrals in \eqref{PI-transverseX}. To this end, note that the non-vacuum term in the exponential, viz, $\sum^{\nroots}_{j=1}\textrm{sgn}(\dot{x}^{\LCp}(t_j))r_{j\LCperp}x^{\LCperp}(t_j)$ is linear in the worldline coordinates, akin to coupling to an extra external source. Hence, the path integral is Gaussian and evaluates to the following compact (and semi-classically exact) form
\begin{align}\label{shock-scalar-propagator1}
    \mathcal{D}^{(s)p'p}_{N}=
    {\int_N}
    \int\!\prod_{j=1}^{\nroots}
    {\hat \ud}^2 r_{j \LCperp}
    W(r_{j\LCperp})\,
    \e^{iS_{\tilde{\mathcal{J}}}[\tilde{x}_{\text{cl}}(\tau);0]}
    \Big\rvert_{\textrm{lin.}\,\varepsilon} \;,
\end{align}
in which $S_{\tilde{\mathcal{J}}}[\tilde{x}_{\text{cl}}(\tau);0]$ is the classical action in vacuum, but where now the classical path $\tilde{x}_{\text{cl}}$ obeys the same boundary conditions as $x_{\text{cl}}$, and the same form of differential equation,  $\ddot{x}=2\tilde{\mathcal{J}}$, but with a {net source} $\tilde{\mathcal{J}}$ defined by
\begin{align}
    \tilde{\mathcal{J}} ^\mu =\mathcal{J}^\mu -\sum_{j=1}^{{\overline N}} \delta(\tau-t_j) \,
    {\tilde k}^\mu_j\;, \qquad \text{with} \qquad {\tilde k}^\mu_j := \textrm{sgn}(\dot{x}^{\LCp}_{\text{cl}}(t_j)) \, r_{j}^\LCperp \delta^\mu_\LCperp \;.
\end{align}
The form of $\tilde{\mathcal{J}}$ suggests that the classical action $S_{\tilde{\mathcal{J}}}[\tilde{x}_{\text{cl}}(\tau);0]$ resembles that in vacuum, with $N+\nroots$ photons, where the additional photons have (purely transverse) momenta ${\tilde k}_j^\mu$. We will now show that this correspondence can be made concrete at the level of off-shell currents, where a suitable auxiliary `polarisation' will appear to complete the current $\tilde{\mathcal{J}}$.

The expression \eqref{shock-scalar-propagator1} for the dressed propagator is deceptively simple in that, modulo the extra $W$ factors, it appears only as complex as the corresponding expression in vacuum. However, a detailed inspection reveals complications: in particular that the roots $t_j\equiv t_j(\mathcal{X})$ \emph{and} their number $\nroots\equiv \nroots(\mathcal{X})$ in the domain $\tau\in(0,T)$, are functions of $\mathcal{X}=(x^{\LCp},p'^{\LCp},\{k_i^{\LCp}\},\{\tau_i\},T)$. This dependence significantly complicates the proper time integrals over $\{\tau_i\}$ and $T$, and the integral over the center of mass mode $x^{\LCp}$, in \eqref{shock-scalar-propagator1}.

\begin{figure}[t!]
\centering
\begin{subfigure}[b]{0.45\textwidth}
\begin{tikzpicture}
    \begin{scope}[decoration={
    markings,
    mark=at position 0.5 with {\arrow{>}}}
    ] 
            \draw[line width=0.4mm,red] (-2,2) -- (2,-2);
            \draw[line width=0.2mm,postaction={decorate}](-0.75,-2) -- (-.25,-1.5);
            \draw (-0.9,-2.2) node {$p$};
            \draw[line width=0.2mm,postaction={decorate}] (-.25,-1.5) -- (0,-.5);
            \draw[line width=0.2mm,postaction={decorate}](0,-.5) -- (.75,.5);
            \draw[line width=0.2mm,postaction={decorate}](.75,.5) -- (1,2);
            \draw (1.2,2.3) node {$p'$};
            \draw[->,draw=blue, snake it] (.25,-2)--(-.25,-1.5);
            \draw (.35,-2.2) node {$k_1$};
            \draw[->,draw=blue, snake it] (0,-.5) -- (-2,1.5);
            \draw (-2.2,1.3) node {$k_2$};
             \draw[draw=red,fill=orange] (.22,-0.22) circle (1.5pt);
             \draw[->,draw=blue, snake it] (2,-.75) -- (.75,.5);
             \draw (2,-1) node {$k_3$};
        \end{scope}
\end{tikzpicture}
\caption{}
  \end{subfigure}  
\begin{subfigure}[b]{0.45\textwidth}
\begin{tikzpicture}
    \begin{scope}[decoration={
    markings,
    mark=at position 0.5 with {\arrow{>}}}
    ] 
            \draw[line width=0.4mm,red] (-2,2) -- (2,-2);
            \draw[line width=0.2mm,postaction={decorate}](0,-2) -- (0,-1);
            \draw[line width=0.2mm,postaction={decorate}] (0,-1) -- (.5,0);
            \draw[line width=0.2mm,postaction={decorate}](.5,0) -- (-.75,.25);
            \draw[line width=0.2mm,postaction={decorate}](-.75,.25) -- (-1,2);
            \draw (-1,2.3) node {$p'$};
            \draw[->,draw=blue, snake it] (0,-1) --(-2,1);
            \draw (0,-2.2) node {$p$};
            \draw[->,draw=blue, snake it] (2,-1.5) -- (.5,0);
            \draw (-2.2,.8) node {$k_1$};
             \draw[draw=red,fill=orange] (.32,-0.32) circle (1.5pt);
             \draw[draw=red,fill=orange] (-.13,.13) circle (1.5pt);
             \draw[draw=red,fill=orange] (-.83,.83) circle (1.5pt);
             \draw[->,draw=blue, snake it] (-.75,.25) -- (-2,1.5);
             \draw (2,-1) node {$k_2$};
              \draw (-2.2,1.5) node {$k_3$};
        \end{scope}
\end{tikzpicture}
\caption{}
\end{subfigure}
\caption{Sketches of example $N=3$ scattering; in (a) and (b) the kinematics obey, respectively violate, the positivity constraint (\ref{positivity}), with the number of roots ${\bar N}=1$ and ${\bar N}=3$ respectively. The sketches are to be interpreted as semi-classical representations of contributions to the scattering process, projected onto the $x^{\LCpm}$-plane. The red line represents the support of the PP-wave, $x^{\LCp}=0$.}
\label{fig:positivity}
  \end{figure}
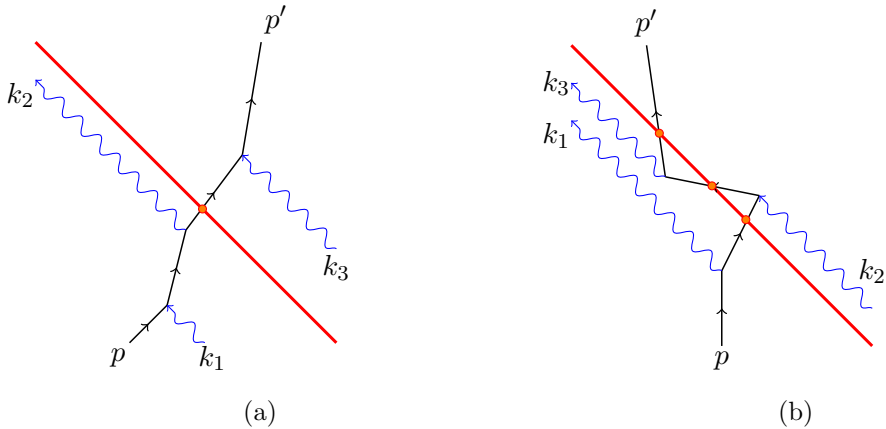  

Our strategy, then, is to make as few reasonable assumptions as possible in order to establish control over the subtleties introduced by the `root structure' of $x_{\text{cl}}^{\LCp}(\tau)$. We make the assumptions
\begin{align}\label{positivity}
    n\cdot p^{\prime} >0 \qquad \text{and} \qquad n\cdot \Big(p^{\prime}+\sum_{i\in\mathcal{U}}k_i\Big) >0  \quad \forall\;\; \mathcal{U}\subseteq \{1,2,3,...,N\}\,,
\end{align}
which we refer to as the `positivity constraint'.  The first of (\ref{positivity}) is not really a restriction, since $n\cdot p^\prime\,, n\cdot p>0$ would, in any case, be enforced when performing LSZ reduction (of the scalar legs) to obtain the off-shell currents, hence we lose nothing here. The real assumption is the second of (\ref{positivity}). This states that the sum of $p^{\prime\LCp}$ and \emph{any} subset of the emitted/absorbed photon momenta remains positive (see Fig.(\ref{fig:positivity}) for a graphical illustration using the $N=3$ case). Note that this \textit{is} automatically satisfied for e.g.~multiple nonlinear Compton scattering in which all photons $k_i$ are emitted, $e^{-}\rightarrow e^{-}+N\gamma$. The positivity constraint is also satisfied for processes in which all photons are \emph{absorbed}, $e^{-}+N\gamma\rightarrow e^{-}$; though the $n\cdot k_i$ in (\ref{positivity}) all become negative in this case, \emph{overall} momentum conservation in the lightfront direction effectively re-imposes the constraint.  For other processes, e.g.~the Compton effect in a background where photons are both absorbed and emitted, (\ref{positivity}) imposes a nontrivial constraint on the allowed kinematic phase space, to which the results below are thus restricted.

The reason for demanding (\ref{positivity}) is that, under this assumption {and the gauge choice $\varepsilon_{i}^{\LCp} = 0$}, $\dot{x}^{\LCp}_{\text{cl}}(\tau)$ becomes positive definite (for any $\{\tau_i\}$), as can be seen from the explicit form
\begin{align}
   \dot{x}^{\LCp}_{\text{cl}}(\tau)=2p'^{\LCp}+ 2\sum_{i=1}^{N}k_i^{\LCp}\Theta(\tau_i-\tau)\,,
\end{align}
and hence ${\bar N}$, the number of roots of $x_{\text{cl}}^{\LCp}(\tau)$, in the interval $0<\tau<T$ is \emph{at most} one. This greatly simplifies (\ref{shock-scalar-propagator1}). In what follows we explore the structure of (\ref{shock-scalar-propagator1}) within this restricted, but physically motivated domain, discussing the restriction a little further below.

When the number of roots is at most 1, we can in particular easily separate the dressed propagator into contributions from the two possible cases $\nroots=0$ and $\nroots=1$, as follows. We can write
\begin{align}\label{shock-scalar-propagator2}
    \mathcal{D}^{(s)p'p}_{N}=&
    {\int_N}
    \e^{iS_{\mathcal{J}}[x_{\text{cl}}(\tau);0]}\bigg[1-\int_{0}^{T}\ud\tau_{\scaleto{(N+1)\mathstrut}{5pt}}\dot{x}^{\LCp}_{\text{cl}}(\tau_{\scaleto{(N+1)\mathstrut}{5pt}})\delta(x_{\text{cl}}^{\LCp}(\tau_{\scaleto{(N+1)\mathstrut}{5pt}}))\bigg]\bigg\rvert_{\textrm{lin.}\,\varepsilon}\\\nonumber
    &+
    {\int_N}
    \int_{0}^{T}\ud\tau_{\scaleto{(N+1)\mathstrut}{5pt}}\!
    {\int {\hat \ud}^2 r_\LCperp}
    W(r_{\perp})\dot{x}^{\LCp}_{\text{cl}}(\tau_{\scaleto{(N+1)\mathstrut}{5pt}})\delta(x_{\text{cl}}^{\LCp}(\tau_{\scaleto{(N+1)\mathstrut}{5pt}}))\,
    \e^{iS_{\tilde{\mathcal{J}}}[\tilde{x}_{\text{cl}}(\tau);0]\big\rvert_{t_1\rightarrow \tau_{\scaleto{(N+1)\mathstrut}{5pt}}}}
    \bigg\rvert_{\textrm{lin.}\,\varepsilon}\!,
\end{align}
where we have introduced additional factors under an integral over a new variable $\tau_{\scaleto{(N+1)\mathstrut}{5pt}}$.
This can of course be trivially performed, setting $\tau_{\scaleto{(N+1)\mathstrut}{5pt}}\to t_1$, the location of the single root of $x^\LCp_{\text{cl}}=0$, if it exists.
As such, $\tilde{\mathcal{J}}$ in the second line is to be understood with the replacement $t_1\rightarrow \tau_{\scaleto{(N+1)\mathstrut}{5pt}}$ before the integral is performed, as indicated. With this, it becomes clear that only the first, respectively second, line of (\ref{shock-scalar-propagator2}) survives in the case that $\nroots=0$, respectively $\nroots=1$, as promised.
This split of the dressed propagator helps reveal interesting features of the corresponding scattering amplitudes, as we will now see.  

\begin{figure}[t]
\centering
\begin{subfigure}[b]{0.45\textwidth}
\begin{tikzpicture}
    \begin{scope}[decoration={
    markings,
    mark=at position 0.5 with {\arrow{>}}}
    ] 
            \draw[line width=0.4mm,red] (-2,2) -- (2,-2);
            \draw (-2.2,2.2) node {$x^{\LCp}\!\!=\!\!0$};
            \draw[line width=0.2mm,postaction={decorate}, densely dotted](0.375,-1.75)--(1.5,.5);
            \draw[line width=0.2mm,postaction={decorate}](1.5,.5) -- (2,1.5);
            \draw[line width=0.2mm,postaction={decorate}] (2,1.5) -- (2.25,2.5);
            \draw[line width=0.2mm,postaction={decorate}](2.25,2.5) -- (2.5,3);
            \draw[->,draw=blue, snake it] (3.5,0) -- (2,1.5);
            \draw[->,draw=blue, snake it] (2.25,2.5) -- (1.25,3.5);
            \draw[line width=0.2mm, dashed,red] (2.5,-.5) -- (-.5,2.5);
             \draw[draw=red,fill=orange] (.8333,-0.8333) circle (1.5pt);
            \draw (2.5,-.8) node {$x^{\LCp}\!\!=\!\!x_{\textrm{cl}}^{\LCp}(0)$};
        \end{scope}
\end{tikzpicture}
\caption{}
  \end{subfigure}  
\begin{subfigure}[b]{0.45\textwidth}
\begin{tikzpicture}
    \begin{scope}[decoration={
    markings,
    mark=at position 0.5 with {\arrow{>}}}
    ] 
            \draw[line width=0.4mm,red] (-2,2) -- (2,-2);
            \draw (-2.2,2.2) node {$x^{\LCp}\!\!=\!\!0$};
             \draw[line width=0.2mm,postaction={decorate}, densely dotted](-.5,-2)--(0,-1);
            \draw[line width=0.2mm,postaction={decorate}](0,-1) -- (.5,0);
            \draw[line width=0.2mm,postaction={decorate}] (.5,0) -- (.75,1);
            \draw[line width=0.2mm,postaction={decorate}](.75,1) -- (1,1.5);
            \draw[->,draw=blue, snake it] (2,-1.5) -- (.5,0);
            \draw[->,draw=blue, snake it] (.75,1) -- (-.25,2);
            \draw[line width=0.2mm, dashed,red] (1,-2) -- (-2,1);
            \draw[draw=red,fill=orange] (.34,-.34) circle (1.5pt);
            \draw (1,-2.3) node {$x^{\LCp}\!\!=\!\!x_{\textrm{cl}}^{\LCp}(0)$};
        \end{scope}
\end{tikzpicture}
\caption{}
\end{subfigure}
\caption{(a) and (b) show representative, contrasting cases of classical worldline solutions $x^{\LCpm}_{\textrm{cl}}(\tau)$ for $N=2$, under the positivity constraint. {Continuous black lines represent $x^{\LCpm}_{\textrm{cl}}(\tau)$ in the range $\tau\in(0,T)$ and as such are relevant for the propagator $\mathcal{D}^{(s)p'p}_{N}$. Dotted lines represent the extension of $x^{\LCpm}_{\textrm{cl}}(\tau)$ to the entire real line $\tau\in(-\infty,\infty)$ and hence are relevant to $\mathcal{A}^{(s)p'p}_{N}$. For $\tau\in(0,T)$ the worldline in (a) does not cross the PP-wave ($\nroots=0$), while that in (b) crosses exactly once ($\nroots=1$). Note, though, that \textit{both} worldlines cross the PP-wave exactly once when extended to the proper-time domain relevant to $\mathcal{A}^{(s)p'p}_{N}$. }}
\label{fig:shock_cross}
\end{figure}
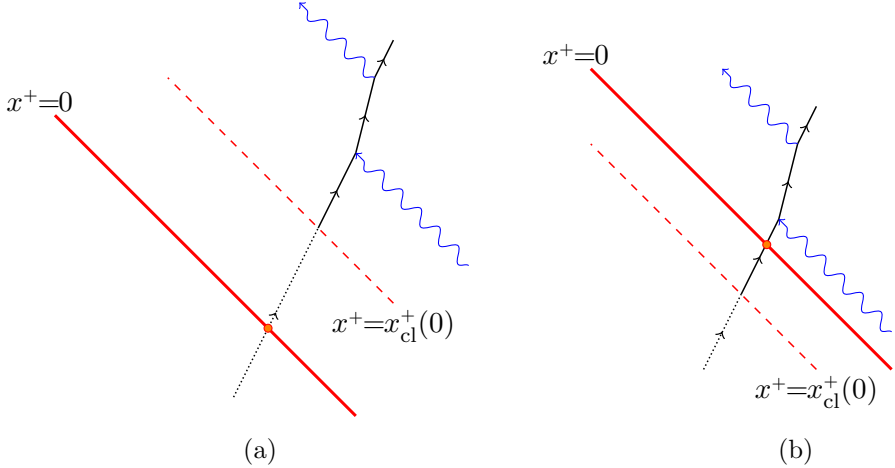

While the relevant domain for $x_{\text{cl}}^{\LCp}(\tau)$ in the master formulae for propagator is $(0,T)$, when applying LSZ reduction to obtain the scattering amplitude, the domain becomes the entire real-line $\mathbb{R}$, as shown, for instance, in \cite{Mogull:2020sak} and \cite{Copinger:2023ctz} -- and summarised above in section \ref{secAmputation}. The number of roots of $x_{\text{cl}}^{\LCp}(\tau)$ in the later domain, under the same assumptions, is exactly one (see, Fig. (\ref{fig:shock_cross})). Therefore, only the second line on the right-hand side of \eqref{shock-scalar-propagator2} survives the LSZ reduction. So, we might as well focus only on that term henceforth, and refer to the same as $\widetilde{\mathcal{D}}^{(s)p'p}_{N}$ for convenience. The expression for $\widetilde{\mathcal{D}}^{(s)p'p}_{N}$, after rewriting $\delta(x_{\text{cl}}^{\LCp}(\tau_{\scaleto{(N+1)\mathstrut}{5pt}}))$ in its Fourier representation is given by
\begin{align}\label{prpagator-vacuum-form0}
   {(-ie)}\widetilde{\mathcal{D}}^{(s)p'p}_{N}=
   {\int_{N+1}}
   \int {{\hat \ud}^3 k_{\scaleto{(N+1)\mathstrut}{5pt}}}
   W(k_{\scaleto{(N+1)\perp\mathstrut}{5pt}})\dot{x}^{\LCp}_{\text{cl}}(\tau_{\scaleto{(N+1)\mathstrut}{5pt}})
   \e^{iS_{\tilde{\mathcal{J}}}[\tilde{x}_{\text{cl}}(\tau);0]+ik_{\scaleto{(N+1)+\mathstrut}{5pt}}x_{\text{cl}}^{\LCp}(\tau_{\scaleto{(N+1)\mathstrut}{5pt}})}\Big\rvert_{\textrm{lin.}\,\varepsilon}\,,
\end{align}
where we have also relabelled $r_{\perp}=k_{\scaleto{(N+1)\perp\mathstrut}{5pt}}$. The above expression invites the identification of an additional auxiliary polarisation, $\varepsilon_{{\scaleto{(N+1)\mathstrut}{5pt}} \mu }=n_{\mu}W(k_{\scaleto{(N+1)\perp\mathstrut}{5pt}})$, so that the term in square-bracket can be written suggestively as $\varepsilon_{\scaleto{(N+1)\mathstrut}{5pt}}\cdot \dot{x}_{\text{cl}}(\tau_{\scaleto{(N+1)\mathstrut}{5pt}})$ and, thus, resembles a worldline vertex operator for photon emission/absorption. With this identification, we introduce an auxiliary four-vector $k_{\mu}$ by declaring $k_{\scaleto{(N+1)\mathstrut}{5pt}-}=0$, and follow the trick (borrowed for the worldline formalism from string theory) of exponentiating the prefactor to obtain
\begin{align}\label{prpagator-vacuum-form00}
   {(-ie)}\widetilde{\mathcal{D}}^{(s)p'p}_{N}=
   {\int_{N+1}}
   \int {{\hat \ud}^3 k_{\scaleto{(N+1)\mathstrut}{5pt}}}
   \e^{iS_{\tilde{\mathcal{J}}}[\tilde{x}_{\text{cl}}(\tau);0]+ik_{\scaleto{(N+1)+\mathstrut}{5pt}}x_{\text{cl}}^{\LCp}(\tau_{\scaleto{(N+1)\mathstrut}{5pt}}) + W(k_{\scaleto{(N+1)\perp\mathstrut}{5pt}})\dot{x}^{\LCp}_{\text{cl}}(\tau_{\scaleto{(N+1)\mathstrut}{5pt}})}\Big\rvert_{\textrm{lin.}\,\varepsilon,\, W}\,.
\end{align}
Two remarks are in order. Firstly, this is analogous to the Bern-Kosower Master Formula derived for QED scattering amplitudes in vacuum. Secondly, since the Master Formula for vacuum scattering amplitudes, (\ref{master-vac}), also arises from a semi-classically exact path integral with current $\mathcal{J}$, it is straightforward to see that we can write $\widetilde{\mathcal{D}}^{(s)p'p}_{N}$ in the compact form
\begin{align}\label{prpagator-vacuum-form}
    (-ie)\widetilde{\mathcal{D}}^{(s)p'p}_{N}=
    \int\! {\hat \ud}^4 k_{\scaleto{(N+1)\mathstrut}{5pt}}
    {\hat \delta}(k_{\scaleto{(N+1)\mathstrut}{5pt}}\cdot n)\,
    D^{p'p}_{N+1}
    \,
    \Big\rvert_{\varepsilon_{\scaleto{(N+1)\mathstrut}{5pt}}
    =
    n_{\mu}W(k_{\scaleto{(N+1)\perp\mathstrut}{5pt}})}\,,
\end{align}
where, $D^{p'p}_{N+1}$ is the photon dressed propagator in vacuum -- see (\ref{master-vac}). It is then straightforward to see that the corresponding amplitude $\mathcal{A}^{(s)p'p}_{N}$ is simply given by
\begin{align}
    (-ie)\mathcal{A}^{(s)p'p}_{N}=
    \int\! {\hat \ud}^4 k_{\scaleto{(N+1)\mathstrut}{5pt}}
    {\hat \delta}(k_{\scaleto{(N+1)\mathstrut}{5pt}}\cdot n)\,
    \mathcal{A}^{p'p}_{N+1}\Big\rvert_{\varepsilon_{\scaleto{(N+1)\mathstrut}{5pt}\mu}
    =
    n_{\mu}W(k_{\scaleto{(N+1)\perp\mathstrut}{5pt}})}\,.
\end{align}
This result says that the $N$-photon off-shell current in an impulsive PP-wave is equal to the $(N+1)$-photon current in vacuum, for a specific choice of $(N+1)^\text{th}$ photon momentum and polarisation. This would be a trivial statement \emph{if} we had treated interactions with the background perturbatively, but we have not -- the background has been treated non-perturbatively. This means the \emph{all-orders} interaction with the PP-wave background, on the left-hand side of (\ref{prpagator-vacuum-form}) can be, recalling that the right-hand side is linear in $W$, `factorised' into a single effective interaction in vacuum.

This is somewhat reminiscent of eikonal exponentiation, in the sense that the $2\to2$ eikonal amplitude is simply the Born amplitude, i.e.~single photon exchange, multiplied by a phase coming from the all-orders resummation of higher loop corrections. The eikonal approximation is of course reproduced by $1\to 1$ scattering on a shockwave background~\cite{tHooft:1987vrq,Jackiw:1991ck}.
What is particularly interesting about (\ref{prpagator-vacuum-form}) is that it holds not just for $1\to 1$ on the background, but for $1\to 1 + N$ (satisfying the positivity constraint); it would be interesting to connect this explicitly to $2\to 2+N$ scattering in the eikonal limit; for a review of related topics see~\cite{DiVecchia:2023frv}. 
    
\subsection{PP-waves as averaged plane waves}\label{sec:scalar-plane}
%
We now show that under the assumption (\ref{positivity}), scattering on a PP-wave background can be factorised in a second, complementary way: it is equivalent to scattering on an impulsive \textit{plane wave} background, where the strength of the plane wave is averaged over at the amplitude level. This result was established for the case of single-photon emission ($N=1$) in~\cite{Adamo:2021jxz}, but here we extend it to all multiplicity.

Recall that in expression (\ref{PI-transverseX}) for the dressed propagator, the background-field dependence was captured entirely by the appropriate number of $U(1)$ factors defined in \eqref{u1-factors}. With this in mind, consider the integral $F^{p'p}_N$ defined by
\be\label{propagator_u1form}
F^{p'p}_{N} 
    =
    \int{\hat \ud}^2 r_{\LCperp} W(r_{\LCperp})
    \int_{N}\!\int_{x}\!\mathcal{D}x(\tau) 
    \bigg[\int_{0}^{T}\!\ud t \, |\dot{x}^{\LCp}(t)|\delta(x^{\LCp}(t))\,
    \e^{i\,\textrm{sgn}(\dot{x}^{\LCp}(t))r_{\LCperp}x^{\LCperp}(t)}\bigg]
    \e^{iS_{\mathcal{J}}[x;0]}\bigg\rvert_{\textrm{lin.}\,\varepsilon}\;.
\ee
This is similar in form to (\ref{propagator_mom_PI_mixed}), except that 1) there is no background field in the action, 2) there is an integral over $W(r_\LCperp)$ outside all other integrals, and 3) we have inserted, in large square brackets, the same `projector' as in (\ref{shock-scalar-propagator2}). We know the the latter reduces to a single U(1) factor, as in (\ref{u1-factors}), when $x^\LCp$ is replaced by its classical expression and the positivity constraint is obeyed. Note that at this stage, though, \emph{none} of the path integrals have been performed; the $x$ which appears is the full path-integral variable.

{We now want to show that $F_N^{p'p}$ is equivalent to $\tilde{\mathcal{D}}^{(s)p'p}_{N}$ at the level of scattering amplitudes (and off-shell currents), \emph{on the support} of the positivity constraint. To see this, observe that the evaluation of the $x^\LCpm$ path integrals in (\ref{propagator_u1form}) proceeds precisely as for both plane waves and shockwaves, and the resulting $x^\LCp_{\text{cl}}$ is the same. Thus, we recover (\ref{shock-scalar-propagator1}) and then, performing the transverse integrals, we would recover the \emph{second} line of (\ref{shock-scalar-propagator2}) with the single insertion of $W$, which we already know is the only piece that survives LSZ truncation. This establishes the equivalence of $F_N^{p'p}$ and $\widetilde{\mathcal{D}}^{(s)p'p}_{N}$ at the amplitude level.}

Now that we see the relevance of $F_N^{p'p}$, we can simplify it further, performing the $\ud t$ integral in (\ref{propagator_u1form}). This appears complicated because the number of roots now depends on the fluctuation, but we know that, once we perform the $x^\LCpm$ integrals, $x^\LCp\to x^\LCp_{\text{cl}}$ which will kill any contributions except that appearing in $\widetilde{D}^{(s)p'p}_{N}$. Thus we can write
\be\label{F-almost-there}
    F_N^{p'p}
     =\int{\hat \ud}^2 r_{\LCperp}
    W(r_{\LCperp})
     \int_{N}\int_{x}\!\mathcal{D}x\, 
    \e^{iS_{\mathcal{J}}[x;0]+i\,\textrm{sgn}(\dot{x}^{\LCp}(t_1))r_{\LCperp}x^{\LCperp}(t_1)}
    \bigg\rvert_{\textrm{lin.}\,\varepsilon} \;,
\ee
Thus, $F_N^{p'p}$ is indeed equivalent to $\tilde{D}^{(s)p'p}_{N}$ under the positivity constraint.
A similar argument means that the exponential in (\ref{F-almost-there}) may be written in the suggestive form
\begin{align}\label{planewave_action}
    S_{\mathcal{J}}[x;0]+\,\textrm{sgn}(\dot{x}^{\LCp}(t_1))r_{\LCperp}x^{\LCperp}(t_1)&=S_{\mathcal{J}}[x;0]+\int_{0}^{T}\,\delta(x^{\LCp}(\tau))\dot{x}^{\LCp}(\tau)r_{\LCperp}x^{\LCperp}(\tau)\,\ud\tau \;,
\end{align}
where the second term is exactly what would appear in the case of a background impulsive plane wave $A^{(\textrm{pw})}$ with potential
\begin{align}\label{def:pw}
    eA^{(\textrm{pw})}_{\mu}=-n_{\mu}\delta(n\cdot x)\, r_\LCperp x^\LCperp \;, 
\end{align}
with $r_\LCperp$ now representing the (instantaneously nonzero and homogeneous) electromagnetic fields of the impulsive wave.
This reveals that the truncated propagator, subject to (\ref{positivity}), is given by `averaging' over the corresponding propagator in a plane wave background over all possible field strengths $r_\LCperp$, with a weight given by $W(r_\LCperp)$:
\begin{align}\label{propagator-pw-form2}
    \tilde{\mathcal{D}}^{(s)p'p}_{N}=\int{\hat \ud}^2 r_{\LCperp}W(r_{\LCperp})
    {\mathcal{D}}^{(\textrm{pw})p'p}_{N}\,.
\end{align}
It immediately follows, performing LSZ reduction on both sides of \eqref{propagator-pw-form2}, that the $2+N$-point scattering amplitudes (and off-shell currents) in an impulsive PP-wave background are given by the same average over the corresponding quantities $\mathcal{A}^{(\textrm{pw})p'p}_{N}$ in plane wave backgrounds:
\begin{align}\label{pw-result}
    \mathcal{A}^{(s)p'p}_{N}=\int\!{\hat \ud}^2r\, W(r_{\perp})\mathcal{A}^{(\textrm{pw})p'p}_{N}\,.
\end{align}
This generalises a result previously found for single photon emission, $e^{-}\rightarrow e^{-}+\gamma$ in PP-wave background~\cite{Adamo:2021jxz} to \emph{arbitrary} multiplicity. We have achieved this with surprisingly little effort, beyond that needed to identify the underlying structure of the $(2+N)$-point amplitudes in plane wave backgrounds previously computed in~\cite{Copinger:2023ctz}.  While this result holds only under the positivity constraint, we repeat that that this is no constraint at all for the processes of e.g.~multiple nonlinear Compton scattering, that is $N\geq 1$ photon \emph{emission}, for which the constraint is automatically obeyed.

\subsubsection*{Discussion}

Given the above, it is worth comparing with the transplanckian scattering of a colour-charged scalar, with the emission of $N$ gluons, in an Aichelburg-Sexl spacetime, as studied in~\cite{Lodone:2009qe}. There it was indeed observed that the amplitude has only one scalar-shock interaction vertex, which in our language translates to $\nroots=1$. What we have seen here is that there is a more general kinematic regime, defined by the positive constraint \eqref{positivity}, in which amplitudes share the same analytic simplicity as the cases of N-photon emission/absorption. {This can be compared to the case of the one-loop calculations of \cite{Tarasov:2019rfp}, where it was found that under the eikonal approximation (virtual) matter loops intersected the shockwave exactly $\nroots = 2$ times.}

We also note that all of the amplitudes (\ref{pw-result}), for each $N$, scale in the same way with increasing field strength, as they are all linear in $W$ (in which all details of the background are contained). This is potentially interesting in the context of higher \emph{loop} corrections in background field perturbation theory. While the loop contributions naively scale with powers of the (small) perturbative coupling, explicit calculations show that the effective expansion parameter actually depends also on the \emph{strong} coupling to the background, and thus higher loop corrections can dominate over lower loop corrections~\cite{RitusRN1,Fedotov:2016afw,Heinzl:2021mji,Mironov:2021ohk,Fedotov:2022ely}; the implication is that background field perturbation theory requires resummation for extremely strong backgrounds. We can see from (\ref{pw-result}) that in the case of PP-wave backgrounds, there are parts of these loops (cuts) which scale quadratically with $W$, \emph{independent} of loop number. Of course, this is only part of the loop parameter space, and if the positivity constraint is not fulfilled it is entirely possible for the tree amplitudes to scale as $W^N$, implying a loop contribution of $W^{2N}$ at $N$ loops. It would be interesting to investigate this further, especially given the known differences in high energy vs.~strong field behaviour of QED loop corrections~\cite{Podszus:2018hnz,Ilderton:2019kqp}.

\section{Fermion scattering on impulsive PP-waves}\label{sec:spinor}
In this section, we extend the results above to the case of \emph{spinor} particles scattering on impulsive PP waves.
The major difference compared to the scalar case is the presence of the Feynman spin factor in our expressions ($\mathfrak{W}$, defined below) which accounts for the spin degrees of freedom and their coupling to the external photons and background. 

\subsection{The kernel and related quantities}

We begin by introducing the relevant spinor quantities. We consider first not the photon-dressed spinor propagator itself (defined below as $\mathcal{S}$), but a closely related object called the spinor kernel $\mathcal{K}$. Not only is this object simpler {than the dressed spinor propagator}, but it more closely resembles the dressed \emph{scalar} propagator which we have already met, although with the addition of a spin factor. It also turns out that $\mathcal{K}$ contains complete information about the currents (with on-shell fermions) for the free and plane wave dressed cases (see~\cite{ Ahmadiniaz:2021gsd,Copinger:2023ctz}), and, as we shall show, the PP-wave case. Explicitly, in momentum space and for an arbitrary background field, the photon-dressed kernel can be written as
\be\label{sp_propagator_mom_PI}
    \mathcal{K}^{p'p}_{N}
    = \int_N
    \int_{x(0)=x}\mathcal{D}x(\tau)\, \e^{iS_{\mathcal{J}}[x(\tau);A]}
    \textrm{Spin}[A](f_{1}; \ldots ; f_{N})
    \Big|_{\textrm{lin.}\,\varepsilon}\;.
\ee
Here $\textrm{Spin}[A]$ encodes the spin degrees of freedom and their coupling to both the background field and the external photons. It is a function of the $f_j^{\mu\nu} := k^{\mu}_j\varepsilon^{\nu}_j-k^{\nu}_j\varepsilon^{\mu}_j$, that is the linearised field strengths associated to photon $j$ of momentum $k_{j}^{\mu}$ and polatisation $\varepsilon_{j}^{\mu}$. $\textrm{Spin}[A]$ itself is
\begin{align}\label{spin-fac-def}
    \textrm{Spin}[A](f_{1}; \ldots ; f_{N}) &= 
    \mathrm{symb}^{-1} \mathfrak{W}_{\eta}[A](f_1; \ldots ;f_N)\\
    \mathfrak{W}_{\eta}[A](f_1; \ldots ;f_N) 
    &\coloneqq 2^{-\frac{D}{2}}\oint_{\mathrm{A/P}}\mathcal{D}\psi(\tau)\, 
    \e^{i\widetilde{S}_{\mathrm{B}}[\psi(\tau),x(\tau),A]}
    \e^{\sum_{j=1}^N \psi_{\eta}(\tau_j) \cdot f_j(x(\tau)) \cdot \psi_{\eta}(\tau_j)}
\end{align}
in which $\mathfrak{W}$ is a path integral over antiperiodic (A/P) Grassmann fields $\psi^{\mu}(\tau)$, and we write $\psi_\eta(\tau)\coloneqq \psi(\tau)+\eta$ for Grassmann $\eta$ for brevity. The spinor worldline action minimally couples these worldline fields to the gauge field:\be
\label{SB-def}
    \widetilde{S}_{\mathrm{B}}[\psi(\tau),x(\tau),A]=\int_{0}^{T}\!\ud\tau\Big[\frac{i}{2}\psi\cdot\dot{\psi}+ie\,\psi_\eta(\tau)\cdot F(x(\tau))\cdot\psi_\eta(\tau)\Big]\,.
\ee
The inverse symbolic map, represented as $\mathrm{symb}^{-1}$ above, acts on the 
Grassmann variables $\eta$ to convert them into Dirac matrices, and thus provide the matrix structure of the kernel -- see~\cite{Copinger:2023ctz} for more details.
We give the relation between the kernel and the propagator below, as we do not need it yet. Note that in (\ref{sp_propagator_mom_PI}) the projection onto multi-linear order in the polarisation vectors is to be achieved by extracting $\varepsilon_{i}$'s from either the orbital or spin parts of the action.

As before, let us record the plane wave and vacuum cases of the spinor kernel as they will be used later. The plane wave case reads, again in the gauge $\varepsilon_i^\LCp=0$,
\begin{align}
    \mathcal{K}^{(\textrm{pw})p'p}_{N} = &(-ie)^{N}\!\int \! \ud^{4}x\!
    \int_{0}^{\infty}\!\ud T\,
    \e^{i({\tilde p}'^2-m^2) T}
    \int_{0}^{T}\prod_{i=1}^{N}\! \ud\tau_i \\ \nonumber
    & \e^{i({\tilde p}'+K-p)\cdot x} \, \e^{ i(2{\tilde p}'+K)\cdot g-i\sum_{i,j=1}^{N}\Big[\frac{1}{2}|\tau_i-\tau_j|k_i\cdot k_j-i\,\,\textrm{sgn}(\tau_i-\tau_j)\varepsilon_i\cdot k_j+\varepsilon_i\cdot\varepsilon_j\delta(\tau_i-\tau_j)\Big]}\\\nonumber
    & \e^{-\int_{0}^{T}(2{\tilde p}' \cdot a(\tau)-a^2(\tau))\ud\tau-2\sum_{i}^{N}\left[\int_{0}^{\tau_i}k_{i}\cdot a(\tau)\ud\tau-i\varepsilon_i\cdot a(\tau_i)\right]}
    \textrm{Spin}[a](f_{1}; \ldots ; f_{N})\bigg|_{\textrm{lin.}\,\varepsilon}\;,
\end{align}
whose spin factor can be exactly evaluated~\cite{Copinger:2023ctz} as
\begin{align}\label{generating_functional}
\mathfrak{W}_{\eta}(f_1; \ldots ;f_N) 
=&
\frac{\delta}{\delta \j_1}\cdot f_1
\cdot\frac{\delta}{\delta \j_1}\cdots
\frac{\delta}{\delta \j_N}\cdot f_N\cdot\frac{\delta}{\delta \j_N}\notag\\
&\e^{-\int_{0}^{T}\!\ud\tau \, [\eta \cdot f(\tau) \cdot \eta +{\j(\tau)} \cdot \eta]
-\int_{0}^{T}\!\ud\tau\,
\ud\tau'\,[\eta \cdot f(\tau) \cdot \mathfrak{G}(\tau, \tau') \cdot \j(\tau')+\frac{1}{4}\j(\tau)\cdot \mathfrak{G}(\tau, \tau')\cdot \j(\tau')]}\Big|_{\j=0}\,,
\end{align}
for plane wave field strength $f$, and with the fermion Green's function (in the plane wave background) being
\be
\mathfrak{G}(\tau,\tau')=\mathrm{sgn}(\tau-\tau')
\Bigl[1-2\int_{\tau'}^{\tau}\!\ud\sigma \,f(\sigma)
+2\Big(\int_{\tau'}^{\tau}\!\ud \sigma\, f(\sigma)\Big)^{2}\Bigr]
+\int_{0}^{T}\!\ud\sigma' \,f(\sigma')\Bigl[1-2\int_{\tau'}^{\tau}\!\ud \sigma\, f(\sigma)\Bigr]\,.
\ee
The vacuum case reads analogously,
\begin{align} \label{eqKNVav}
    K^{p'p}_{N} = &(-ie)^{N}{\hat\delta}^4(p'+K-p)
    \int_{0}^{\infty}\!\ud T\, \e^{i(p'^2-m^2) T}
    \int_{0}^{T}\prod_{i=1}^{N}\! \ud\tau_i \\ \nonumber
    & \e^{ i(2p'+K)\cdot g-i\sum_{i,j=1}^{N}\Big[\frac{1}{2}|\tau_i-\tau_j|k_i\cdot k_j-i\,\,\textrm{sgn}(\tau_i-\tau_j)\varepsilon_i\cdot k_j+\varepsilon_i\cdot\varepsilon_j\delta(\tau_i-\tau_j)\Big]}
    \textrm{Spin}[0](f_{1}; \ldots ; f_{N})
    \Big|_{\textrm{lin.}\,\varepsilon}\;,
\end{align}
where the spin factor is just (\ref{generating_functional}) evaluated at zero (plane wave) field strength, see~\cite{Ahmadiniaz:2020wlm}.

Regarding LSZ reduction, the operations listed in~(i-iii,i-iv) for the (vacuum, plane wave) cases extend to the spinor amplitudes in both backgrounds, supplemented with~\cite{Ahmadiniaz:2021gsd, Copinger:2023ctz} (iv) sandwich the spin factor between free spinors $\bar{u}_{s'}(p')$ and $u_s(p)$, and (v) multiply by $1/(2m)$. This leads to the following expression for the off-shell current for the vacuum case as
\begin{align}
    \mathcal{M}_{Ns's}^{p'p}
    =&(-ie)^N \hat{\delta}^4({p'+K-p})
    \int_{-\infty}^{\infty}\prod_{i=1}^{N} \ud\tau_i\, 
    \delta\bigg(\sum_{j=1}^N\frac{\tau_j}{N}\bigg)
     \frac{1}{2m}\bar{u}_{s'}(p')\mathrm{Spin}[0](f_{1}; \ldots  f_{N})u_{s}(p)
    \nonumber\\
    &\e^{i({{p}' +p})\cdot g-i\sum_{i,j=1}^{N}\bigl(\frac{1}{2}|\tau_i-\tau_j|k_i \cdot k_j-i\,\mathrm{sgn}(\tau_i-\tau_j)\varepsilon_i \cdot k_j+\delta(\tau_i-\tau_j)\varepsilon_i \cdot \varepsilon_j\bigr)}\Big|_{\textrm{lin.}\,\varepsilon}\;,
\end{align} 
and for the plane wave case as
\begin{align}\label{spinor_pw_amplitude}
    \mathcal{M}_{Ns's}^{(\text{pw})p'p}
    =& (-ie)^N\hat{\delta}_{\perp,-}({\tilde p}'+K-p)\int_{-\infty}^{\infty}\!\ud x^{\LCp}e^{i(K+p'-p)_\LCp x^{\LCp}}
    \int_{-\infty}^{\infty}\prod_{i=1}^{N} \ud\tau_i\, 
    \delta\bigg(\sum_{j=1}^N\frac{\tau_j}{N}\bigg)
    \nonumber\\
    & \e^{-i\int_{-\infty}^{0}\bigl[2\tilde{p}' \cdot a(\tau)-a^{2}(\tau)\bigr]d\tau-i\int_{0}^{\infty}\bigl[2p'\cdot\delta a(\tau)-\delta a^{2}(\tau)\bigr]\ud\tau-2i\sum_{i=1}^{N}\bigl[\int_{-\infty}^{\tau_i}k_{i} \cdot a(\tau)\ud\tau-i\varepsilon_i\cdot a(\tau_i)\bigr]}\nonumber \\
    & \e^{i({\tilde p}' +p)\cdot g-i\sum_{i,j=1}^{N}\bigl(\frac{1}{2}|\tau_i-\tau_j|k_i \cdot k_j-i\,\mathrm{sgn}(\tau_i-\tau_j)\varepsilon_i \cdot k_j+\delta(\tau_i-\tau_j)\varepsilon_i \cdot \varepsilon_j\bigr)}
    \nonumber \\
    & \frac{1}{2m}\bar{u}_{s'}(p')\mathrm{Spin}[a](f_{1}; \ldots ; f_{N})u_{s}(p)
    \Big|_{\textrm{lin.}\,\varepsilon}\;,
\end{align} 
for which the exponent of the spin factor reads
\be
    -\int_{-\infty}^{\infty}\!\ud\tau \, [\eta\cdot f\cdot \eta + {\j}\cdot \eta]-\int_{-\infty}^{\infty}\!\ud\tau\int_{-\infty}^{\infty}\ud\tau'\bigl[\eta\cdot f(\tau)\cdot \mathfrak{G}(\tau,\tau')\cdot \j(\tau')+\tfrac{1}{4}\j(\tau)\cdot\mathfrak{G}(\tau,\tau')\cdot\j(\tau')\bigr]\,.
\ee
We remark that generally LSZ reduction acts on the full propagator, and not the kernel; however, for the free~\cite{Ahmadiniaz:2021gsd} and plane wave~\cite{Copinger:2023ctz} cases above only the kernel remains after reduction. We describe this step in more detail for the impulsive PP-wave background below.
%

\subsection{{Effective vacuum current}}
We now specialise to impulsive PP-wave backgrounds and show, as for the scalar case, that the all-orders interaction of a massive particle with an impulsive PP-wave, also dressed by the $N$-photon off-shell current, is equivalent to an effective single photon interaction in vacuum.

Using the field strength (\ref{F-J-PP}), the action (\ref{SB-def}) may be written
\be\label{spinor-WL-action-PPwave}
    \widetilde{S}_{\mathrm{B}}^{(s)}[\psi(\tau),x(\tau),A]
    =
    \int_{0}^{T}\!\ud\tau\Big[\frac{i}{2}\psi\cdot\dot{\psi}+2ie\delta(x^{\LCp})\psi_\eta^{\LCp}(\tau)\psi_\eta^\LCperp(\tau)\partial_\LCperp\Phi(x^{\LCperp})\Big]\,.
\ee
To compute the $x$-integral in (\ref{spin-fac-def}) we again expand about the vacuum classical solution, $x(\tau)=x_{\text{cl}}(\tau)+q(\tau)$, with $x_{\text{cl}}(\tau)$ as in Sec.~\ref{sec:scalar}. The action (\ref{spinor-WL-action-PPwave}) evaluated on $x^{\pm}_{\text{cl}}$ is 
\be
    \widetilde{S}_{\mathrm{B}}^{(s)}[\psi(\tau), x^{\pm}_{\rm{cl}}(\tau), x^{\perp}(\tau),A]
    =
    \frac{i}{2}\int_{0}^{T}\!\ud \tau\,\psi\cdot\dot{\psi}
    +
    2ie\sum_{j}^{\nroots}
    \frac{1}{|\dot{x}_{\text{cl}}^{\LCp}(t_{j})|}\psi_\eta^\LCp(t_{j})
    \psi_\eta^\LCperp(t_{j})
     {\partial_\LCperp \Phi(x^{\LCperp}(t_{j}))}
    \,.
\ee
Since each term is composed of bilinears in Grassmann variables we may write the exponential of the action as a product over roots:
\be\label{414}
    \e^{i\widetilde{S}_{\mathrm{B}}^{(s)}[\psi(\tau),x(\tau),a]}
    \longrightarrow
    \e^{\frac{i}{2}\int_{0}^{T}\!\ud\tau\psi\cdot\dot{\psi}}\prod_{j}^{\nroots}
    \e^{-2e\frac{1}{|\dot{x}_{\text{cl}}^{\LCp}(t_{j})|}\psi_\eta^\LCp(t_{j})
    \psi_\eta^\LCperp(t_{j})
   {\partial_\LCperp
    \Phi(x^{\LCperp}(t_{j}))}\,.}
\ee
Expanding out the exponentials for each root $j$, only terms to linear order survive, as all higher terms contain vanishing products of Grassmann variables $\psi_\eta^\LCp(t_{j})^{2}=0$. Thus (\ref{414}) becomes
\be \label{eqDPerpPhi}
    \e^{i\widetilde{S}_{\mathrm{B}}^{(s)}[\psi(\tau),x(\tau),a]}
    \longrightarrow
    \e^{\frac{i}{2}\int_{0}^{T}\!\ud\tau\psi\cdot\dot{\psi}}\prod_{j}^{\nroots}
    \Bigl[
    1-2e\frac{1}{|\dot{x}_{\text{cl}}^{\LCp}(t_{j})|}\psi_\eta^\LCp(t_{j})
    \psi_\eta^\LCperp(t_{j})
    {\partial_\LCperp
    \Phi(x^{\LCperp}(t_{j}))
    \Bigr]\,.}
\ee
{A similar observation was made in~\cite{Tarasov:2019rfp}  -- there as here it is} due to the delta-function support of the background, which localises the worldline parameter integral on discrete values of proper time: in a general background one would find products of $\psi(\tau)$ at \emph{different} times, and thus the higher order terms would survive. Now combining (\ref{eqDPerpPhi}) with the scalar contribution after integrating out $\delta x^{\LCpm}$, i.e.~equation (\ref{U1-product}), we can generate the $\partial_{\perp} \Phi$ in (\ref{eqDPerpPhi}) by differentiating the scalar exponent, allowing us to write
\begin{align}\label{e-product-e}
    &\e^{i\widetilde{S}_{\mathrm{B}}^{(s)}[\psi(\tau),x(\tau),a]}\prod_{j}^{\nroots}
    \e^{ie\,\mathrm{sgn}(\dot{x}_{\text{cl}}^{\LCp}(t_{j}))\Phi(x^{\LCperp}(t_{j}))}\notag\\
    &\longrightarrow \e^{\frac{i}{2}\int_{0}^{T}\!\ud\tau\psi\cdot\dot{\psi}}
    \prod_{j}^{\nroots}
    \Bigl[
    1+\frac{2i}{\dot{x}_{\text{cl}}^{\LCp}(t_{j})}\psi_\eta^\LCp(t_{j})
    \psi_\eta^\LCperp(t_{j})
    {\partial_\LCperp}
    \Bigr]
    \e^{ie\,\mathrm{sgn}(\dot{x}_{\text{cl}}^{\LCp}(t_{j}))\Phi(x^{\LCperp}(t_{j}))}\,.
\end{align}
Introducing the Fourier transform of the U(1) factors of $\Phi$ as in~\eqref{u1-factors}, the term in square brackets in (\ref{e-product-e}) becomes
\be\label{shock_spin}
    {\int \hat{\ud}^{2}r_{j\perp} \Big[1-\frac{2}{|\dot{x}_{\text{cl}}^{\LCp}(t_{j})|}\psi_\eta^\LCp(t_{j})
    \psi_\eta^\LCperp(t_{j})r_{j\LCperp}\Big] W(r_{j\perp}) \e^{i\, \mathrm{sgn}(x^{+}) r_{j\perp} x^{\perp}}}\;,
\ee
and we see that as for the scalar propagator, under the Fourier integral, the path integral has been linearised.

Retracing our steps to the worldline action in an impulsive PP-wave background 
under the Fourier integral, we find that we may replace the spinor action with the related functional
\be
    \widetilde{S}_{\mathrm{B}}^{(s)}[\psi(\tau),x(\tau),\tilde{a}]=\int_{0}^{T}\!\ud\tau\Big[\frac{i}{2}\psi\cdot\dot{\psi}+i\,\psi_\eta(\tau)\cdot \tilde{f}\cdot\psi_\eta(\tau)\Big]\,,
\ee
where {$\tilde{f}_{\mu\nu}=\sum_{j=1}^{\bar{N}} |\dot{x}_{\text{cl}}^{\LCp}(t_{j})|^{-1} \delta(\tau - t_j)[n_{\mu}\delta_{\nu}^\LCperp-\delta _{\mu}^\LCperp n_{\nu}] r_{j \LCperp}$} is an effective field strength tensor.
Therefore we may express the spinor kernel analogous to the scalar propagator,~\eqref{shock-scalar-propagator1}, as 
\be
    \mathcal{K}^{(s)p'p}_{N} = 
    \int_N
    \,
    \int\prod_{i=1}^{\nroots}
    \hat{\ud}^2 r_{j\LCperp}W_{j}(r_{j\LCperp})\e^{iS_{\tilde{\mathcal{J}}}[\tilde{x}_{\text{cl}}(\tau);0] }\,
    \mathrm{symb}^{-1}\mathfrak{W}_{\eta}[\tilde{f}](f_1; \ldots ;f_N) 
    \Big\rvert_{\textrm{lin.}\,\varepsilon}\,,
\ee
where $\mathfrak{W}_{\eta}[\tilde{f}](f_1; \ldots ;f_N)$ is~\eqref{spin-fac-def}, but with $eF\to \tilde{f}$.

To make progress, we again adopt the positivity constraint (\ref{positivity}) in precisely the same form as for the scalar case. Let us then write, as before, the part of the spinor kernel that will ultimately remain after the LSZ reduction ({we discuss the spinor-specific details of this reduction in more detail below}). As for the scalar, this is the $\nroots=1$ contribution which can be projected onto by introducing an auxiliary integral, c.f.,~\eqref{prpagator-vacuum-form0},
\begin{align}
    (-ie)\widetilde{\mathcal{K}}^{(s)p'p}_{N} =&\int_{N+1}
    \int{\hat{\ud}}^3 k_{\scaleto{(N+1)\mathstrut}{5pt}}
    \e^{iS_{\tilde{\mathcal{J}}}[\tilde{x}_{\text{cl}}(\tau);0]+ik_{\scaleto{(N+1)\LCp\mathstrut}{5pt}} x^\LCp_{\text{cl}}(\tau_{\scaleto{(N+1)\mathstrut}{5pt}})}\notag\\
    &\bigl[W(k_{\scaleto{(N+1)\perp\mathstrut}{5pt}})\dot{x}^\LCp_{\text{cl}}(\tau_{\scaleto{(N+1)\mathstrut}{5pt}})\bigr]
    \mathrm{symb}^{-1}\mathfrak{W}_{\eta}[\tilde{f}](f_1; \ldots ;f_N) 
    \Big{|}_{\textrm{lin.}\,\varepsilon}\,.
\end{align}
Examining the terms in the brackets coupled to the background dependent part of the spin factor,~\eqref{shock_spin} as it appears above, we find a prefactor
\be
     W(k_{\scaleto{(N+1)\perp\mathstrut}{5pt}})\dot{x}^\LCp_{\text{cl}}(\tau_{\scaleto{(N+1)\mathstrut}{5pt}})-2W(k_{\scaleto{(N+1)\perp\mathstrut}{5pt}})\psi_\eta^\LCp(\tau_{\scaleto{(N+1)\mathstrut}{5pt}})
    \psi_\eta(\tau_{\scaleto{(N+1)\mathstrut}{5pt}})\cdot k_{\scaleto{(N+1)\mathstrut}{5pt}}\,,
\ee
{so that $\e^{i\widetilde{S}_{\mathrm{B}}^{(s)}[\psi(\tau),x(\tau),a]}    \e^{ie\Phi(x^{\LCperp}(t_{j}))}$ is replaced by
\begin{equation}
    \int \hat{\ud}^{3}k_{\scaleto{(N+1)\mathstrut}{5pt}} \int_{0}^{T} \ud \tau_{\scaleto{(N+1)\perp\mathstrut}{5pt}}\Big[  W(k_{\scaleto{(N+1)\perp\mathstrut}{5pt}})\dot{x}^\LCp_{\text{cl}}(\tau_{\scaleto{(N+1)\mathstrut}{5pt}})-2W(k_{\scaleto{(N+1)\perp\mathstrut}{5pt}})\psi_\eta^\LCp(\tau_{\scaleto{(N+1)\mathstrut}{5pt}})
    \psi_\eta(\tau_{\scaleto{(N+1)\mathstrut}{5pt}})\cdot k_{\scaleto{(N+1)\mathstrut}{5pt}}\Big] \e^{i  k_{\scaleto{(N+1)\mathstrut}{5pt}} \cdot x}\,,
\end{equation}
which contains exactly} the spinor vertex operator for emission/absorption of an auxiliary, off-shell photon with momentum $k_{\scaleto{(N+1)\mathstrut}{5pt}\mu}$ with the identification of $\varepsilon_{\mu\scaleto{(N+1)\mathstrut}{5pt}}=n_{\mu}W(k_{\scaleto{(N+1)\mathstrut}{5pt}})$. Thus one may immediately write down for the spinor kernel, {in analogy to (\ref{prpagator-vacuum-form}),}
\be\label{prpagator-vacuum-form-spinor-new-label}
    (-ie)\widetilde{\mathcal{K}}^{(s)p'p}_{N}
    =
    \int\!    {\hat \ud}^4 k_{\scaleto{(N+1)\mathstrut}{5pt}} \, 
    {\hat{\delta}}(k_{\scaleto{(N+1)\mathstrut}{5pt}}\cdot n)K^{p'p}_{N+1}\Big\rvert_{\varepsilon_{\mu\scaleto{(N+1)\mathstrut}{5pt}}\rightarrow n_{\mu}W(k_{\scaleto{(N+1)\perp\mathstrut}{5pt}})}\,,
\ee
where $K_{N+1}^{p'p}$ is the photon-dressed kernel in vacuum \cite{Ahmadiniaz:2020wlm}.

\subsubsection{From $\mathcal{K}_{N}$ to scattering amplitudes}
Although the above shows that the kernel behaves similarly to the scalar propagator, the connection to off-shell currents and scattering amplitude must now be made. {In particular, the complete \textit{correlation function}, respectively $N$-photon-dressed \textit{correlation function}, in position space are related to the kernel through (see e.g.~\cite{Ahmadiniaz:2021gsd,Copinger:2023ctz})}
\begin{align}
    \mathcal{S}^{x'x} &= \big(-i\slashed{\partial}_{x'} + e \slashed{A}^{(s)}(x') - m\big)\mathcal{K}^{x'x} \;,\\
 \mathcal{S}^{x'x}_N &=  
    (-i\slashed{\partial}_{x'}+e\slashed{ A}^{(s)}(x^{\prime })-m)    {\mathcal K}_{N}^{x'x}+e\sum_{i = 1}^{N}\slashed{\varepsilon}_{i} \e^{i k_{i} \cdot x'}  {\mathcal K}_{N-1}^{x'x}\,.\label{S-K-2}
\end{align}
The first term on the RHS of (\ref{S-K-2}), involving $\mathcal{K}_{N}$, is called the ``leading term'' with all $N$ external photons drawn from the kernel $\mathcal{K}_{N}$, while the second term is referred to as the ``subleading term'' in which one of the $N$ photons is drawn from the covariant derivative in $\mathcal{S}$ and the remaining $N-1$ are drawn from $\mathcal{K}_{N-1}$. The issue to be addressed here is the possibility that either the subleading term, or the gauge field entering the covariant derivative, in the $N$-photon-dressed correlation function could obstruct the immediate extension of (\ref{prpagator-vacuum-form-spinor-new-label}) to 
the $N$-photon off-shell current since both terms should be considered in the LSZ reduction.

It has been shown in vacuum and plane wave backgrounds that the subleading terms do not contribute to off-shell currents (as they miss the required mass-shell poles to survive LSZ reduction), and that the background field-dependent part of the covariant derivative also drops out during the LSZ procedure (as the external legs are drawn to infinity)  \cite{Ahmadiniaz:2021gsd,Copinger:2023ctz}. The asymptotic behaviour of the impulsive PP wave is similar to the plane wave case studied in~\cite{Copinger:2023ctz}, albeit with vanishing asymptotic gauge ($A(x) \rightarrow 0$ as $x^\LCp \to \pm \infty$ holds for the massive external legs considered here due it only having support at $x^\LCp=0$). Thus, we will be able to show that similar conclusions may be drawn here as well.

To make this more precise, we re-express the LSZ prescription for the fermion scattering amplitude,
\begin{equation}
        \label{eqFermionLSZ}
   \mathcal{M}_{N}^{p'p} = i{\lim_{p'^2,p^2\rightarrow m^2}} \int \!\ud^{4}x'\ud^{4}x\,
    \e^{ip'\cdot x'-ip\cdot x}
    \bar{u}(p')(i\slashed{\partial}_{x'}-m)\mathcal{S}_{N}^{x'x}(-i\overleftarrow{\slashed{\partial}}\!_{x}-m)u(p)\,,
    \end{equation}
{($\bar{u}(p')$ and $u(p)$ are free spinors relevant for the asymptotic fermions)
in terms of the $N$-photon kernel as}
\begin{align}
         \mathcal{M}_{N}^{p'p} = \frac{i}{2m}
    \lim_{p'^2,p^2\rightarrow m^2}
    \int \!
    \ud^{4}x' \ud^{4}x
    \,
    \e^{ip'\cdot x'-ip\cdot x}&\bar{u}(p')(p^{\prime 2} - m^{2})\Bigl\{
    (-1
    {+ \frac{e}{2m}\slashed{A}^{(s)}(x^{\prime \LCp})}
    )\,\mathcal{K}_N^{x'x} \nonumber \\
    &+\frac{e}{2m}\sum_{i=1}^{N}\slashed{\varepsilon}_{i} \e^{ik_{i}\cdot x'}\,\mathcal{K}_{N-1}^{x'x}\Bigr\}(p^2-m^2)u(p)\,.
    \end{align}
Computing the Fourier integrals is straightforward if we exploit 
(\ref{prpagator-vacuum-form-spinor-new-label})
, {for then we immediately obtain} the corresponding relation in position space:
\be
    (-ie)\widetilde{\mathcal{K}}^{(s)x'x}_{N}
    =
    \int\!    {\hat \ud}^4 k_{\scaleto{(N+1)\mathstrut}{5pt}} \, 
    {\hat{\delta}}(k_{\scaleto{(N+1)\mathstrut}{5pt}}\cdot n)K^{x'x}_{N+1}\Big\rvert_{\varepsilon_{\mu\scaleto{(N+1)\mathstrut}{5pt}}\rightarrow n_{\mu}W(k_{\scaleto{(N+1)\perp\mathstrut}{5pt}})}\,.
\ee
Then, using the explicit form of $K_{N+1}^{x'x}$ in (\ref{eqKNVav}) and switching to lightfront coordinates, the integrals over $x^{-}$ and $x^{\prime -}$ yield $\delta$-functions:
\begin{equation}
    \int\! \ud x^{\LCm} \ud x^{\prime\LCm} \, \e^{i p^{\prime} \cdot x^{\prime} - ip\cdot x} K_{N+1}^{x'x} \sim  {{\hat \delta}\big(p^{\prime +} + \sum_{i= 1}^{N+1} k^\LCp_{i} - p^\LCp\big)} \hat{\delta}\big(x^{\prime +} - x^{\LCp} - 2Tp^{\prime \LCp} - 2 \sum_{i=1}^{N+1} k_{i}^{\LCp}\tau_{i}  \big) \;.
\end{equation}
The first delta function expresses momentum conservation, while the second replaces
\begin{equation}
    x^{\prime \LCp} \longrightarrow x^{\LCp} + 2Tp^{\prime \LCp} + 2 \sum_{i=1}^{N+1}  k_{i}^{\LCp}\tau_{i} \,,
\end{equation}
{which, in particular, is applied to the argument of $\slashed{A}^{(s)}(x^{\prime \LCp})$.}
Finally, we apply the steps outlined in Sec.~\ref{secAmputation} which projects onto the $T \rightarrow \infty$ limit of the proper time integrand, where in particular the argument of  $\slashed{A}^{(s)}$ is taken to infinity, where it has no support. This shows that the gauge-potential part of the covariant derivative does not contribute to the on-shell
scattering amplitude in the shockwave. The subleading terms are similarly treated by carrying out the Fourier transform using the explicit form of the position space propagator; the external photon wavefunctions multiplying the kernel now shift the outgoing momentum, $p^\prime \rightarrow p^{\prime} - k_{i}$. Truncation of external legs then produces a contribution proportional to $(p^2-m^2)/((p-k_{i})^2-m^2)$ which vanishes on-shell. 

Therefore, as in plane wave backgrounds, the on-shell amplitude for spinors under the positivity constraint is entirely contained in the leading term in the kernel. This allows us to lift (\ref{prpagator-vacuum-form-spinor-new-label}) to the full scattering amplitude, $\mathcal{M}^{(s)p'p}_{N}$. By implementing LSZ reduction on the LHS of (\ref{prpagator-vacuum-form-spinor-new-label}), we obtain
\be
    (-ie)\mathcal{M}^{(s)p'p}_{N}= 
    \int\!    {\hat \ud}^4 k_{\scaleto{(N+1)\mathstrut}{5pt}} \, 
    \hat{\delta}
    (k_{\scaleto{(N+1)\mathstrut}{5pt}}\cdot n)\mathcal{M}^{p'p}_{N+1}\Big\rvert_{\varepsilon_{\mu\scaleto{(N+1)\mathstrut}{5pt}}\rightarrow n_{\mu}W(k_{\scaleto{(N+1)\perp\mathstrut}{5pt}})}\,,
\ee
where $\mathcal{M}^{p'p}_{N+1}$ is the $(N+1)$-photon vacuum amplitude~\cite{Ahmadiniaz:2020wlm}, in which the $(N+1)^{\rm{th}}$ photon has momentum $k_{\scaleto{(N+1)\mathstrut}{5pt}}$ and polarisation $\varepsilon_{\mu\scaleto{(N+1)\mathstrut}{5pt}}= n_{\mu}W(k_{\scaleto{(N+1)\perp\mathstrut}{5pt}})$.

\subsection{PP-waves as averaged plane waves}

In this section we {generalise the result of Sec.~\ref{sec:scalar-plane} to the spinor case, showing that the $N$-photon amplitudes on the PP-wave background can be expressed as averages of the corresponding amplitudes on plane wave backgrounds. We might expect this to hold given both the scalar results above and the fact that it holds for $N=1$~\cite{Adamo:2021jxz}; indeed the arguments closely parallel those for the scalar theory, so we will be brief.}

In analogy to $F^{p'p}_N$ given in~\eqref{propagator_u1form}, we begin with with a Fourier integral over $W(r_\LCperp)$ outside a full path integral over $x$, projected onto the $\bar{N} = 1$ contribution:
\begin{align}\label{eqGNp}
G^{p'p}_{N} 
    =&
    \int{\hat \ud}^2 r_{\LCperp} W(r_{\LCperp})
    \int_{N}\!\int_{x}\!\mathcal{D}x 
    \bigg[\int_{0}^{T}\!\ud t \, |\dot{x}^{\LCp}(t)|\delta(x^{\LCp}(t))\,
    \e^{i\,\textrm{sgn}(\dot{x}^{\LCp}(t))r_{\LCperp}x^{\LCperp}(t)}\bigg]
    \e^{iS_{\mathcal{J}}[x;0]}\\
    &\mathrm{symb}^{-1}2^{-\frac{D}{2}}\oint_{\mathrm{A/P}}\mathcal{D}\psi(\tau)\, 
    \e^{-\frac{2}{|\dot{x}_{\text{cl}}^{\LCp}(t)|}\psi_\eta^\LCp(t) \psi_\eta^\LCperp(t)r_{\LCperp}+i\widetilde{S}_{\mathrm{B}}[\psi(\tau),x(\tau),0]}
    {\e^{\sum_{j=1}^N \psi_{\eta}(\tau_j) \cdot f_j(x(\tau)) \cdot \psi_{\eta}(\tau_j)}}
    \Big\rvert_{\textrm{lin.}\,\varepsilon}\;.\notag
\end{align}
Then, assuming the positivity constraint, it follows that once the semi-classically exact $x^\LCpm$ integrals are performed, generating the \emph{same} classical solution $x_{\text{cl}}^\LCp$, 
$G^{p'p}_N$ {is found} to be equivalent to  $\widetilde{\mathcal{K}}^{(s)p'p}_N$.
This implies that the 
first term in the exponent in the second line of (\ref{eqGNp}) -- which represents the spin degrees of freedom interacting with the PP-wave -- may be expressed as
\be
    \frac{2}{|\dot{x}_{\text{cl}}^{\LCp}(t_1)|}\psi_\eta^\LCp(t_1)
    \psi_\eta^\LCperp(t_1)r_{1\LCperp}=
    e\int_{0}^{T}\!\ud\tau\,\psi_\eta(\tau)\cdot F^{(\text{pw})}\cdot\psi_\eta(\tau)\,,
\ee
where $F^{(\text{pw})}$ represents the field strength of~\eqref{def:pw}.
Thus, $\widetilde{\mathcal{K}}^{(s)p'p}_N$ may also be interpreted as an average over all possible field strengths $r_\LCperp$ with weight $W(r_\LCperp)$ in a plane wave background:
\be
    \widetilde{\mathcal{K}}^{(s)p'p}_{N}=\int{\hat \ud}^2 r_{\LCperp}W(r_{\LCperp})
    {\mathcal{K}}^{(\textrm{pw})p'p}_{N}\,.
\ee
Since {the impulsive plane wave~\eqref{def:pw} and the impulsive PP-wave~\eqref{PP-wave-def} have the same asymptotic behaviour as} $x^\LCp{\to \pm \infty}$, it is straightforward to see that LSZ reduction will project onto the leading contribution from $\widetilde{\mathcal{K}}^{(s)p'p}_{N}$ and discard the contribution from the gauge potential in the covariant derivative. Hence we may also lift this relation to the on-shell amplitude, so that it may also be expressed an an averaging over a plane wave background:
\be
    \mathcal{M}^{(s)p'p}_{N}=\int\!{\hat \ud}^2r\, W(r_{\perp})\mathcal{M}^{(\textrm{pw})p'p}_{N}\,.
\ee
{Thus, for the $N$-photon scattering amplitudes of both scalar and spinor particles, the PP-wave background acts as a `stochastic' plane wave (under the positivity constraint). This generalises the results of~\cite{Adamo:2021jxz} to arbitrary multiplicity.}

\section{Conclusions}\label{sect:conclusions}
In this paper we have explored tree-level amplitudes and off-shell currents describing {arbitrary}-multiplicity photon emission/absorption from massive charged particles scattering on general impulsive PP wave backgrounds. The coupling of the particles to the background was treated exactly throughout.

Using the worldline formalism, we observed that there exists an interesting (and hitherto not \emph{explicitly} explored) kinematic regime, defined by the positivity constraint (\ref{positivity}), in which the amplitude and off-shell current, in a sense, `factorise'. The factorisation structure can be represented in two distinct forms--- one that relates the photon-dressed propagators to known off-shell currents in vacuum, the other that expresses the same propagator in a related plane wave background.

In the first representation, the $N$-photon off-shell current in the PP-wave background is written as a weighted average of $(N+1)$-photon off-shell currents in vacuum. The average is taken over the (off-shell) momentum of the $(N+1)^\text{th}$ photon, with a specific polarisation determined by the strength and propagation direction of the PP-wave background; the weight is given by a delta function supported on the net change in lightfront momentum. Physically, this picture implies that the all-orders interaction with the PP-wave can be described as a single interaction with an off-shell photon in vacuum. In the second representation, the $N$-photon amplitudes and off-shell currents in the PP-wave background are written as a weighted average of the corresponding quantities in an impulsive plane wave background, with the average now taken over all possible \emph{strengths} of the plane wave, with weight again determined by the PP-wave.

It is notable that we have been able to expose these structures \emph{without} having to evaluate all the worldline parameter integrals explicitly (i.e.~it was not necessary to obtain explicit expressions for the amplitudes in the three scenarios above). Rather, our results came about from exposing an effective Gaussianity in the worldline path integral that rendered it semi-classical exact, a property shared with both the vacuum and plane wave cases.  In this respect, our work builds upon~\cite{Edwards:2017bte, Ahmadiniaz:2017rrk} which also use manipulations under the path integral, thereby allowing functional relationships to be exposed between contributions to the quantum effective action in a homogeneous background.

The fact that our results can be established at arbitrary multiplicity is an additional benefit of the worldline approach, which allows us to consider all numbers of emission/absorption events in a single Master Formula -- such general manipulations would seem difficult to access in the standard formalism of QFT. 

We comment on some possible directions to take in future work. In light of the well-known correspondence between QFT in a shockwave background and ultra-relativistic scattering, our results suggest the existence of 
interesting factorisation structures for $N+4$-point amplitudes in vacuum (two matter lines and $N$ external photon lines) -- it would be interesting to see how the conventional diagrammatic approach reveals such a regime. For a worldline formulation of this problem in vacuum, see~\cite{Bastianelli:2014bfa}.

All possible kinematics for $N$-photon emission-only, and $N$-photon absorption-only, processes are allowed by the positivity constraint. For other processes, one may ask whether the amplitudes evaluated at general kinematics possess e.g.~a series expansion in which the factorisation above appears as the leading order term, and whether more complicated relations between corresponding processes in vacuum or associated plane waves exist.

It would also be interesting to see whether the structures found here extend to loop corrections. Loop diagrams in backgrounds modelling laser fields, for example, are relevant for upcoming vacuum polarisation experiments~\cite{King:2015tba,Karbstein:2019dxo,Gies:2022pla,Borysov:2022cwc,Macleod:2023asi}, and for the Ritus-Narozhny conjecture~\cite{Fedotov:2016afw,Mironov:2021ohk,DiPiazza:2020kze,Heinzl:2021mji}. Background-field calculations beyond one-loop are extremely challenging, but some progress has recently been made, in a variety of backgrounds, through the use of novel methods, see e.g.~\cite{Gies:2016yaa,Karbstein:2019wmj,Adamo:2020syc,Dunne:2021acr,Torgrimsson:2021wcj,Podszus:2021lms,Torgrimsson:2021zob,Podszus:2022jia,Torgrimsson:2022ndq}. Higher loop calculations in PP-wave backgrounds, in particular, seem to be largely unexplored, at least in QED (although one can infer something of their behaviour via their cuts, see~\cite{Adamo:2021jxz}). Given the wealth of results obtained from the worldline formalism for homogeneous background fields \cite{Ahmadiniaz:2023vrk, Huet:2020awq, Huet:2018ksz, Dunne:2004xk} and the apparent simplifications it can provide in the case of light-by-light (e.g. four-photon) scattering \cite{Ahmadiniaz:2020jgo, Ahmadiniaz:2023vrk, Ahmadiniaz:2022yam}, it would be interesting to see if our formalism could offer new insights here.

\begin{acknowledgments}
The authors are supported by the EPSRC Standard Grants  EP/X02413X/1 (PC, JPE) and EP/X024199/1 (AI, KR), and the STFC consolidator grant ``Particle Theory at the Higgs Centre,'' ST/X000494/1 (AI).

\end{acknowledgments}
\appendix

\bibliography{BeyondBib}
\bibliographystyle{JHEP}

\end{document}